\documentclass[a4paper,fleqn,usenatbib]{mnras}

\usepackage{newtxtext,newtxmath}
\usepackage[T1]{fontenc}
\usepackage{ae,aecompl}
\usepackage{graphicx}	% Including figure files
\usepackage{amsmath}	% Advanced maths commands
\usepackage{amssymb}	% Extra maths symbols

\usepackage[table,xcdraw]{xcolor}
\usepackage{multirow}
\usepackage{comment} 
\usepackage{booktabs,fixltx2e}
\usepackage[flushleft]{threeparttable}
\usepackage{longtable}
\usepackage{lscape}
\usepackage{xargs} 
\usepackage{caption}
\usepackage[colorinlistoftodos,prependcaption,textsize=tiny]{todonotes}
\usepackage[bottom]{footmisc}

\let\s\iffalse % Copy definition of \iffalse into \s

\def\am{$^{\prime}$ }

\def\hii{\textsc{Hii} region }
\def\hiino{\textsc{Hii} region}
\def\hiis{\textsc{Hii} regions }
\def\hiisno{\textsc{Hii} regions}

\newcommandx{\change}[2][1=]{\todo[linecolor=blue,backgroundcolor=blue!25,bordercolor=blue,#1]{#2}}

\title[Galactic synchrotron distribution]
{Galactic synchrotron distribution derived from 152 H{\sc{ii}} region absorption features in the full GLEAM survey}
\author[H. Su et al.]{
H.~Su$^{1,2,3}$\thanks{E-mail: hongquan.su@icrar.org},
J.~P.~Macquart$^{2}$\thanks{E-mail: J.Macquart@curtin.edu.au},
N.~Hurley-Walker$^{2}$\thanks{E-mail: nhw@icrar.org},
N.~M.~McClure-Griffiths$^{4}$,
\newauthor
C.~A.~Jackson$^{5}$,
S.~J.~Tingay$^{2}$,
W.~W.~Tian$^{1,3,6}$,
B.~M.~Gaensler$^{7,8,9}$,
B.~McKinley$^{2,8}$,
\newauthor
A.~D.~Kapi\'{n}ska$^{10,11}$,
L.~Hindson$^{12}$,
P.~Hancock$^{2}$,
R.~B.~Wayth$^{2}$,
L.~Staveley-Smith$^{10,8}$,
\newauthor
J.~Morgan$^{2}$,
M.~Johnston-Hollitt$^{2}$,
E.~Lenc$^{7,8}$,
M.~E.~Bell$^{13,14,8}$,
J.~R.~Callingham$^{5}$,
\newauthor
K.~S.~Dwarkanath$^{15}$,
B.-Q.~For$^{10}$,
A.~R.~Offringa$^{5}$,
P.~Procopio$^{16}$,
C.~Wu$^{10}$
and
Q.~Zheng$^{17}$
\\
$^{1}$~Key Laboratory of Optical Astronomy, National Astronomical Observatories, Chinese Academy of Sciences, Beijing 100012, China\\
$^{2}$~International Centre for Radio Astronomy Research, Curtin University, Bentley, WA 6102, Australia\\
$^{3}$~University of Chinese Academy of Science, 19A Yuquan Road, Beijing 100049, China\\
$^{4}$~Research School of Astronomy and Astrophysics, Australian National University, Canberra, ACT 2611, Australia\\
$^{5}$~ASTRON, The Netherlands Institute for Radio Astronomy, Postbus 2, 7990 AA, Dwingeloo, The Netherlands\\
$^{6}$~Department of Physics \& Astronomy, University of Calgary, Calgary, Alberta T2N 1N4, Canada\\
$^{7}$~Sydney Institute for Astronomy, School of Physics, The University of Sydney, NSW 2006, Australia\\
$^{8}$~ARC Centre of Excellence for All-sky Astrophysics (CAASTRO), Australia\\
$^{9}$~Dunlap Institute for Astronomy and Astrophysics, University of Toronto, 50 St.\ George Street, Toronto, ON M5S 3H4, Canada\\
$^{10}$~International Centre for Radio Astronomy Research, University of Western Australia, Crawley, WA 6009, Australia\\
$^{11}$~National Radio Astronomy Observatory, Socorro NM 87801, USA\\
$^{12}$~Centre for Astrophysics Research, School of Physics, Astronomy and Mathematics, University of Hertfordshire, College Lane,\\ \,\,\,\,\,\,Hatfield AL10 9AB, UK\\
$^{13}$~University of Technology Sydney, 15 Broadway, Ultimo NSW 2007, Australia\\
$^{14}$~CSIRO Astronomy and Space Science (CASS), PO Box  76, Epping, NSW 1710, Australia\\
$^{15}$~Raman Research Institute, Bangalore 560080, India\\
$^{16}$~School of Physics, The University of Melbourne, Parkville, VIC 3010, Australia\\
$^{17}$~School of Engineering \& Computer Science, Victoria University of Wellington, PO Box 600, Wellington 6140, New Zealand
}
\date{Accepted 2018 June 22. Received 2018 June 22; in original form 2018 April 30}
\pubyear{2018}
\begin{document}
\label{firstpage}
\pagerange{\pageref{firstpage}--\pageref{lastpage}}
\maketitle

\begin{abstract}
We derive the synchrotron distribution in the Milky Way disk from \hii absorption observations over $-40^\circ < l  < 40^\circ$ at six frequencies of 76.2, 83.8, 91.5, 99.2, 106.9, and 114.6~MHz with the GaLactic and Extragalactic All-sky Murchison widefield array survey (GLEAM). 
We develop a new method of emissivity calculation by taking advantage of the Haslam et al., (1981) map and known spectral indices, which enable us to simultaneously derive the emissivity and the optical depth of \hiis at each frequency. We show our derived synchrotron emissivities based on 152 absorption features of \hiis using both the method previously adopted in the literature and our improved method. We derive the synchrotron emissivity from \hiis to the Galactic edge along the line of sight and, for the first time, derive the emissivity from \hiis to the Sun. These results provide direct information on the distribution of the Galactic magnetic field and cosmic-ray electrons for future modelling. 
\end{abstract}


%%%%%%%%%%%%%%%%%%%%%%%%%%%%%%
%%%%%%%%% BODY OF PAPER %%%%%%%%%%

\section{Introduction}

At frequencies from about 10~MHz to 1~GHz, the diffuse emission in the Milky Way is dominated by the synchrotron emission originating from cosmic ray electrons spiralling in the Galactic magnetic field. 
The two-dimensional distribution of this emission has been mapped and used for building the Global Sky Models (e.g. \citealt{Oliveira-Costa2008MNRAS.388..247D, Zheng2017MNRAS.464.3486Z} and references therein). 
However, its three-dimensional distribution is difficult to infer \citep{Beuermann1985A&A...153...17B}, largely due to the difficulty of separating different components along the line of sight. One method of obtaining depth information relies on the presence of optically thick \hiis embedded in this medium. At low radio frequencies near 100~MHz, some \hiis become optically thick to the background synchrotron emission. The absorption of these \hiis enables us to separate the synchrotron emission into components in front of and behind these regions. Using this \hii absorption technique, \citet{Nord2006AJ....132..242N} derived the brightness temperature behind 42 \hiisno, mainly in the northern sky, using data obtained with the 74\,MHz receivers on the Very Large Array (VLA). More recently, \citet{Su2017MNRAS.465.3163S, Su2017MNRAS.472..828S} derived brightness temperatures behind 47 \hiis and detected 306~\hiis in total \citep{Hindson2016PASA...33...20H} using data at 88 MHz from the Murchison Widefield Array (MWA; \citealt{Tingay2013PASA...30....7T,Bowman2013PASA...30...31B} ).

Observations of the synchrotron emissivity obtained in conjunction with \hiis can, in principle, constrain the structure of the Milky Way.  The synchrotron emission distribution is believed to be correlated with the spiral arm structure of the Milky Way, however, there is no firm observational evidence available.  The warp of the Milky Way's disk should also affect the synchrotron distribution. The outer disk warps upwards (northwards) in the first and second quadrants, downwards on the opposite side \citep{Burke1957AJ.....62...90B, Kerr1957AJ.....62...93K}, and at least 12 \hiis  exist in the outer Scutum-Centaurus arm with a distance of about 15~kpc to us \citep{Armentrout2017ApJ...841..121A}. However, a denser sampling of the synchrotron emission distribution is needed to investigate its relationship to such structures.

To date, the distribution of the Galactic synchrotron emission along the line of sight is too sparsely sampled to constrain its complex distribution. Models of the synchrotron emission based on the derived emissivity from \hii absorption is rudimentary \citep{Nord2006AJ....132..242N, Su2017MNRAS.465.3163S}, with the emission usually assumed to be confined to an axisymmetric cylinder with a radius of 20~kpc and a height of 2~kpc. This radius is a reasonable assumption because the most distant \hiis detected so far have Galactocentric radii more than 19 kpc \citep{Anderson2015ApJS..221...26A}, which may present the outer limit to the extent of the massive star-forming disk. The extragalactic synchrotron emission outside of this disk is usually neglected, assumed to be small compared to the disk contribution. 

The purpose of this paper is to present synchrotron emission measurements using low-frequency MWA data to derive the free-free opacities of 152 \hiis in the Milky Way and determine the synchrotron emission in front of and behind these clouds at six frequencies of 76.2, 83.8, 91.5, 99.2, 106.9, and 114.6~MHz. The data from the MWA enable us to triple the sample of \hiino-absorbed measurements from the multi-frequency observations with much-improved angular resolution and surface brightness sensitivity. 
Furthermore, we develop an improved method and re-derive results from other work using this methodology.

In Section~\ref{sec:data} we introduce the data used for this work. The new method we developed is discussed in Section~\ref{sec:method}. In section~\ref{sec:result}, we present our newly-derived emissivities and in Section \ref{sec:discussion} we discuss our results and compare them to previous work. We summarise our findings in Section~\ref{sec:sum}.

\section{Data}
\label{sec:data}

We use data obtained by the MWA as part of the GaLactic and Extragalactic All-sky MWA survey (GLEAM, \citealt{Wayth2015PASA...32...25W}). The data in this work was collected in four weeks within the first year of the GLEAM survey between 2013 June and 2014 July. This survey covers all the sky south of declination +30$^\circ$ corresponding to a Galactic longitude range of $-50^\circ < l  < 60^\circ$ at $b = 0^\circ$ with \hii absorption found in the range of $-40^\circ < l  < 40^\circ$, $-2^\circ < b < 4^\circ$.
\citet{Hurley-Walker2017MNRAS.464.1146H} presented the calibration, imaging and mosaicking of the GLEAM survey, particularly for the extragalactic catalogue. The data reduction of the Galactic plane region will be reported in Hurley-Walker et al. (in prep). 
Here we only highlight that a multiscale clean in WSCLEAN \citep{Offringa2014MNRAS.444..606O} is performed to better deconvolve the complex structures on the Galactic plane. 

The GLEAM survey has an angular resolution of about 4\am at 100~MHz and excellent $u$-$v$ coverage. This resolution is a 30-fold improvement over existing full-sky maps at comparable frequencies, which have angular resolutions $\geq 2^\circ$. This angular resolution enables us to resolve 10\% of the 8000 \hiis in the Wide-Field Infrared Survey Explorer (WISE) \hii catalogue \citep{Anderson2014ApJS..212....1A}.
The angular resolution varies between 5.41\am and 2.89\am depending on the frequency. We convert the average surface brightness of our selected regions to brightness temperature using the listed conversion factors in Table~\ref{tab:gleam_para}. 
Typical root-mean-squared (rms) values of the GLEAM maps are 0.2~Jy~beam$^{-1}$ at 76.2 MHz to 0.1~Jy~beam$^{-1}$ at 114.6 MHz, estimated using the Background and Noise Estimator (BANE) v1.4.6 from the AEGEAN package \citep{Hancock2012MNRAS.422.1812H, Hancock2018PASA...35...11H}.
The GLEAM survey observes across the frequency range between 72 and 231~MHz, but here we utilise data at the lowest six frequencies from 72 to 118 MHz with a bandwidth of 7.68\,MHz each (see Table~\ref{tab:gleam_para}), these being the most pertinent to the detection and characterisation of the absorption features caused by \hiisno.

We use the all-sky 408 MHz map of \citet{Haslam1981A&A...100..209H, Haslam1982A&AS...47....1H} reprocessed by \citep{Remazeilles2015MNRAS.451.4311R} to estimate the total power of Galactic synchrotron emission along the line of sight towards the \hiis at the GLEAM frequencies. The Haslam map is a combination of four different surveys from the Jodrell Bank MkI, Bonn 100 meter, Parkes 64 meter and Jodrell Bank MkIA telescopes. This map is dominated by the Galactic synchrotron emission with 6\% free-free emission \citep{Dickinson2003MNRAS.341..369D} as neglectable contamination for this work. We also neglect the free-free absorption due to the unresolved \hiis and the warm interstellar medium. The reprocessed Haslam map removed the strong point sources in the destriped/desourced (dsds) version. Thus, the contamination of the extra-galactic sources is minimized. 

\begin{table}
\caption{Parameters of the GLEAM survey data with a bandwidth of 7.68 MHz each. The resolution element is described by the beam major axis (BMAJ) and beam minor axis (BMIN).}
\label{tab:gleam_para}
\begin{tabular}{cccc}
\hline
Frequency & BMAJ & BMIN & Conversion factor\tabularnewline
MHz & arcmin & arcmin & Jy beam$^{-1}$ to K\tabularnewline
\hline 
76.2 & 5.41 & 4.43 & 2445.01\tabularnewline
83.8 & 4.78 & 3.89 & 2598.84\tabularnewline
91.5 & 4.35 & 3.54 & 2633.47\tabularnewline
99.2 & 4.03 & 3.30 & 2596.70\tabularnewline
106.9 & 3.99 & 3.22 & 2310.85\tabularnewline
114.6 & 3.63 & 2.89 & 2467.46\tabularnewline
\hline
\end{tabular}
\end{table}

\section{Improved method of emissivity calculation}
\label{sec:method}
A simplified method of calculating the Galactic synchrotron emissivity was adopted by \citet{Nord2006AJ....132..242N} and slightly modified to include the contribution of the measured background by \citet{Su2017MNRAS.465.3163S, Su2017MNRAS.472..828S}. We believe this approach can be improved in two ways.
 
Firstly, it assumes the optical depths of \hiis are much larger than 1. However, this assumption may not be correct for some \hiis because they show only mild absorption ($\tau \sim 1$) at the frequencies used to separate the foreground and background emission. 

\begin{figure}
\includegraphics[width=0.5\textwidth]{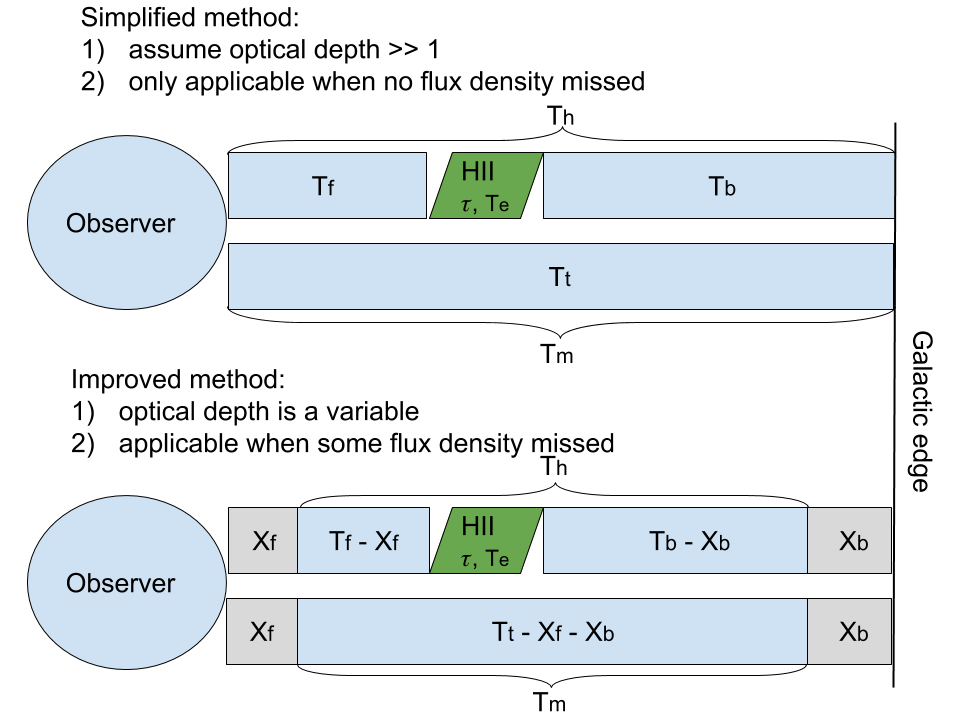}
\caption{A schematic of how the missing flux density affects the derived emissivities in the simplified method (top) and in the improved method (bottom).}
\label{fig:sim_com}
\end{figure}

Secondly, the method underestimates the emissivity behind \hiis when some flux density is resolved out by an interferometric observation, especially when the all-sky ``zero-spacing" component is omitted (see Fig.~\ref{fig:sim_com}). The shortest spacing of the MWA tiles is about 7~metres, corresponding to an angular scale of about 30~degrees, indicating that the MWA is sensitive to the whole sky emission with fluctuations on scales smaller than 30~degrees. Structures on larger angular scales are resolved out by the MWA. This undetected emission has the effect that the derived emissivities are under-estimated. The surface brightness of the Galactic synchrotron emission increases towards lower frequencies, making its contribution large at the $\approx $100\,MHz frequencies relevant to the detection of \hiis compared to 408\,MHz at which the Haslam map was obtained.

To improve this method, we have developed a procedure that attempts to solve for both the optical depth of the \hiis and the brightness temperature of the emission associated with the missing interferometric spacings. We scale the 408~MHz all-sky image to the frequency of interest by a global brightness temperature spectral index ($\alpha$: $S_\nu \propto \nu^\alpha$) to estimate the total power along the line of sight and then use it to deduce the brightness temperature on scales resolved out by our interferometer. We use two data sets from the GLEAM survey with each one containing three frequencies (76.2, 83.8, 91.5~MHz; and 99.2, 106.9, 114.6~MHz ) to perform calculations and assume that synchrotron and optical depths have a power law scaling with the frequency. More details of this new method are described in what follows.

\subsection{Definition of parameters}
\label{sec:def}
Fig.~\ref{fig:sim_com} shows a schematic of the absorption process, indicating the variables needed to solve for the emissivity. As usual, we assume the Galactic synchrotron emission is confined to an axisymmetric cylinder with a radius of 20~kpc and a height of several~kpc. Note that this assumption is only for the definition of the emissivity in Section~\ref{subsec:emi}. We can avoid making this assumption if we are only interested in the brightness temperature instead of the emissivity. 

The measured or known parameters are
the measured brightness temperature in the direction of the absorbed region $T_{\rm h}$,
the measured brightness temperature from the Sun to the Galactic edge in the absence of \hii emission $T_{\rm m}$ (i.e. as derived from a region near the line of sight to the \hiino),
the observation frequency $\nu$, 
the spectral index of the synchrotron brightness temperature $\alpha$, 
the spectral index of the \hii optical depth $\beta$,
the total brightness temperature (without missing flux density) from the Sun to the Galactic edge in the absence of \hii absorption $T_{\rm t}$, and 
the electron temperature of the \hii $T_{\rm e}$. 
$\alpha$ is taken to be $-2.7~\pm~0.1$ for the Milky Way \citep{Guzman2011A&A...525A.138G, Zheng2017MNRAS.464.3486Z}. Note that this spectral index varies between -2.1 and -2.7 depending on the sky regions \citep{Guzman2011A&A...525A.138G}. We use a low spectral index of -2.7 for the synchrotron emission in this work because the high spectral index is due to the thermal free-free absorption of both the \hiis and warm interstellar medium.
$\beta$ is taken to be -2.1 for frequencies $\nu \ll 10^{10}~T_{\rm e}$ ($\nu$ is in GHz and $T_{\rm e}$ is in K) and $T_{\rm e} < 9 \times 10^5$~K derived on page 47 of \citet{Lang1980afcp.book.....L}, which is always true for \hiis at the GLEAM frequencies. Note that $\beta$ is a constant does not mean the \hii must be optically thick; it can be optically thin. The errors caused by these two spectral indices are discussed in Section~\ref{subsec:err}.
$T_{\rm t}$ is derived from the improved Haslam map \citep{Remazeilles2015MNRAS.451.4311R}, scaled from 408~MHz to the GLEAM frequencies using the spectral index of $\alpha$. 
$T_{\rm e}$ is from \citet{Balser2015ApJ...806..199B, Hou2014A&A...569A.125H} and references therein.

The unknown variables are the optical depth of the \hii $\tau$, the total (the sum of the measured and missed) brightness temperature of the synchrotron emission from the \hii to the Galactic edge along the line of sight $T_{\rm b}$, and the corresponding brightness temperature of the synchrotron emission from the \hii to the Sun $T_{\rm f}$, the brightness temperature of the emission on the missing short interferometric spacings respectively between an \hii and the Sun $X_{\rm f}$, and between the Galactic edge and the Sun $X_{\rm b}$. 

The selection criteria of the absorbed region and its nearby background region are the same as those in \citet{Su2017MNRAS.465.3163S}. We define these regions at the lowest frequency of 76.2 MHz and then apply them to all other five frequencies to get the brightness temperatures within these regions. \hiis overlapped with supernova remnants are not selected (e.g. \hii G35.6$-$0.5 with distance measured by \citealt{Zhu2013ApJ...775...95Z}). Note that our selected background regions are about one degree away from the absorbed regions, the supernova remnants in \citet{Green2014BASI...42...47G} catalogue, and obvious point-like sources in the GLEAM survey. Therefore, the contamination of these sources is negligible, although the Haslam map has a low angular resolution of 51$^{\prime}$. 

\subsection{Equations to solve for the optical depth and brightness temperature}
\label{sec:equ}

A single-dish observation can recover the total power along the line of sight in the case that the \hii fills the beam. The brightness temperature is a result of the contributions from three components: the electron temperature of the \hiino, and the brightness temperature of the synchrotron emission behind and in front of the \hii \citep{Kassim1987PhDT........10K},
\begin{equation}
\label{eq:abs}
T_{\rm h} = T_{\rm e}(1-e^{-\tau}) + T_{\rm b}e^{-\tau} + T_{\rm f}.
\end{equation}

An interferometer observation does not sample the large angular scale structures corresponding to visibility measurements at small $u$-$v$ distances. Thus Equation~\ref{eq:abs} should be revised by subtracting the missing term from the brightness temperature both behind and in front of the \hiino,
\begin{equation}
T_{\rm h} = T_{\rm e}(1-e^{-\tau}) + (T_{\rm b}-X_{\rm b})e^{-\tau} + T_{\rm f} - X_{\rm f}.
\end{equation}
Note that this equation does not require the $u$-$v$ coverage to be identical at different frequencies because we do not assume the brightness temperature of the missing term follows the same spectral index. We allow the value of the X terms to float with frequency, as X depends on the angular scale at which emission is being missed, which varies with frequency.

The total brightness temperature from the Sun to the Galactic edge in the absence of \hii absorption is simply the sum of the brightness temperatures behind and in front of the \hiino,
\begin{equation}
\centering
T_{\rm t} = T_{\rm f} + T_{\rm b}. 
\end{equation}

The measured brightness temperature on the source-free region (i.e. immediately adjacent to the \hiino) becomes the difference between the total brightness temperature and that of the brightness temperatures of the emission associated with the missing interferometric $u$-$v$ spacings,
\begin{equation}
T_{\rm m} = T_{\rm t} - X_{\rm f} - X_{\rm b}.
\end{equation}

As well as the above three relations, we have supplementary information that encodes the scaling of the brightness temperature and the optical depth with frequency. The total brightness temperature both behind and in front of the \hii should follow a power law distribution,
\begin{equation}
\begin{aligned}
& T_{\rm b} \propto \nu^{\alpha}, \\ 
& T_{\rm f} \propto \nu^{\alpha}, \\
& \tau \propto \nu^{\beta}. \\
\end{aligned}
\end{equation}

We apply Equations (2)-(5) to our measurements at different frequencies to solve for the optical depth of \hiis and the brightness temperatures behind and in front of each \hiino. In summary, we have:

\begin{equation}
\label{eq:com}
\begin{aligned}
& T_{\rm h_i} = T_{\rm e}(1-e^{-\tau_{\rm i}}) + (T_{\rm b_i}-X_{\rm b_i})e^{-\tau_{\rm i}} + T_{\rm f_i} - X_{\rm f_i}, \\
& T_{\rm t_i} = T_{\rm f_i} + T_{\rm b_i}, \\
& T_{\rm m_i} = T_{\rm t_i} - X_{\rm f_i} - X_{\rm b_i},  \\
& T_{\rm b_i} = T_{\rm b_1} \left( \frac{\nu_{\rm i}}{\nu_1} \right)^{\alpha}, \\
& T_{\rm f_i} = T_{\rm f_1} \left( \frac{\nu_{\rm i}}{\nu_1} \right)^{\alpha}, \\
& \tau_{\rm i} = \tau_{1} \left( \frac{\nu_{\rm i}}{ \nu_1} \right)^{\beta}, \\
\end{aligned}
\end{equation}
where the subscript $i=(1,2,3)$ indexes the frequencies from low to high. A minimum of three frequencies is required to solve for the unknown variables. 

We derive the value of $\tau$, $T_{\rm b}$, $T_{\rm f}$, $X_{\rm b}$, $X_{\rm f}$ using Equation~\ref{eq:com}. Using two sets of three frequencies data, we obtain emissivities at the six frequencies listed in Table~\ref{tab:gleam_para}. We use data at 76.2, 83.8, and 91.5~MHz to derive the emissivites at these three frequencies. And then use another three frequencies of 99.2, 106.9, and 114.6~MHz to perform the same analysis. So we derive the emissivities at six different frequencies. Table~\ref{tab:emi} lists the emissivities at 76.2 MHz only. We did not use other combinations of data to derive emissivities. We can derive $X_{\rm b}$ and $X_{\rm f}$ using the Haslam map to estimate the total emission along the line of sight. We then compare this total emission with that measured. Therefore, our equations can find out how much emission is undetected in our observations. 

\subsubsection{Definition of emissivity}
\label{subsec:emi}
The emissivity is defined to be the brightness temperature divided by the corresponding distance, i.e.,
\begin{equation}
\begin{aligned}
& \epsilon_{\rm b} = T_{\rm b} / D_{\rm b}, \\
& \epsilon_{\rm f} = T_{\rm f} / D_{\rm f},
\end{aligned}
\end{equation}
where 
$\epsilon_{\rm b}$ is the average emissivity between the \hii and the Galactic edge, 
$\epsilon_{\rm f}$ is the average emissivity between the \hii and the Sun,
$D_{\rm b}$ is the distance from the \hii to the Galactic edge, and
$D_{\rm f}$ is the distance from the \hii to the Sun. The value of $D_{\rm f}$ is  derived from \citet{Anderson2014ApJS..212....1A, Anderson2017arXiv171007397A, Balser2015ApJ...806..199B, Hou2014A&A...569A.125H}. The value of $D_{\rm b}$ is calculated from $D_{\rm f}$ assuming a Galactocentric radius of 20~kpc.

\section{Results}
\label{sec:result}
\subsection{Emissivities from simplified equations}

We calculate the synchrotron emissivities behind \hiis using the 152 \hii absorption features detected in the GLEAM survey using the previous simplified method (Col. 11 in Table~\ref{tab:emi}). The last column in Table~\ref{tab:emi} shows the emissivities derived from the simplified method described in \citet{Su2017MNRAS.465.3163S}. The measurements are made at six frequencies from 76.2 to 114.6 MHz. The emissivities behind \hiis at 76.2 MHz are plotted in Fig.~\ref{fig:comp_com_sim_glon}. The derived emissivities in the fourth Galactic quadrant are consistent with our previous results in \citet{Su2017MNRAS.465.3163S, Su2017MNRAS.472..828S}. The emissivities in the first quadrant are consistent with those in \citet{Nord2006AJ....132..242N} within three standard deviations. 

We check the spectral index of the emissivities at six frequencies derived from each \hiino. The average index is about -1.5, which is higher than the expected synchrotron emission spectral index of -2.7 (see Fig.~\ref{fig:si_sim}). The difference between these observed two spectral indexes is most likely caused by the missing flux density mentioned in Section~\ref{sec:method}. To produce a flat spectrum with a spectral index of -1.5, the brightness temperature of the emission on scales that are resolved out should be frequency-dependent, with brightness temperatures under-estimated at lower frequencies in our observations, even though our lower frequencies recover more of the extended emission than the high frequencies: this demonstrates that we need to improve this simplified method to derive more accurate emissivities.

\subsection{Emissivities derived from our new method}
Using the improved method described in Section~\ref{sec:method}, we obtained the synchrotron emissivities and \hii optical depths at six frequencies simultaneously (see Table~\ref{tab:emi}). Fig.~\ref{fig:emi76_on_mw_img} shows the emissivities at 76.2~MHz and the paths over which these emissivities are averaged. The electronic version of the full tables with our derived emissivities at all six frequencies is available from VizieR. Figs.~\ref{fig:emib76_glon_glat} and \ref{fig:emibf76} show our derived emissivities at 76.2 MHz both behind and in front of \hiis using the improved method. 

\subsection{Error estimation}
\label{subsec:err}
For the emissivities derived from simplified equations, we propagate the error throughout the simple equations to estimate their errors. For our improved method, the equations are too complex to permit directly calculating the uncertainty of each solution caused by the variance of the known parameters from the measurements. The sources of the error include 
\begin{itemize}
\item the error of the \hii electron temperature,
\item the error of the distance from \hii to us,
\item the rms of the brightness temperature for the absorbed region in the GLEAM map,
\item the rms of the brightness temperature for the background region in the GLEAM map,
\item the rms of the brightness temperature for the background region in the Haslam map,
\item the variation of the spectral indices of the synchrotron brightness temperature and the \hii optical depth.
\end{itemize}
We use a Monte Carlo method to statistically estimate the error of these solutions caused by the first five error sources. Specifically, we use the values of known parameters to calculate the solutions and then sample around these parameters. We set each input parameter to be a random number following a Gaussian distribution with a mean from the best input value and a standard derivation from our one-sigma measurement error. 
Using these new input parameters, we can find new solutions. By repeating the calculation, we get a distribution of each solution and then calculate the one-sigma upper and lower limits. The estimated errors are about 10-90\% of the emissivity values (see Table~\ref{tab:emi}). Note that we do not include the contribution of the spectral indices of the synchrotron brightness temperature and the \hii optical depth because finding the solutions becomes computationally expensive with these two spectral indices included. The spectral index of the brightness temperature cause a difference of about 15\% of the emissivity values, estimated from the variance of the Haslam map scaling, when this spectral index changes from -2.7 to -2.6. Although this causes extra error to the derived emissivities, it is still necessary to use the Haslam map; otherwise, the derived emissivities behind the \hiis will be under-estimated due to the missing flux density, and the emissivities in front of \hiis cannot be calculated. The error contributed by the spectral index of the \hii optical depth is small ($<$~1\% of the emissivity) because the term $e^{-\tau}$ in Equation~\ref{eq:com} is small when the optical depth is much larger than one.

We check which input parameter dominates the errors of the final emissivities. We set only one input parameter to be a random number while setting all other parameters to be constants. Then, similarly to the above error estimation, we calculate the one-sigma upper and lower limits of the emissivities. The error contribution of each input parameter is shown in Fig.~\ref{fig:err_from_each_input}. We find the rms of the brightness temperature of the Haslam map contributes the most to the final errors of the derived emissivities. 

In the future, new maps using new data processing techniques may be able to recover the total power along the line of sight, which will avoid extrapolating the Halam map from 408 MHz to the GLEAM frequencies. For example, \citet{Eastwood2017arXiv171100466E} use a new widefield imaging technique, named the Tikhonov-regularized m-mode analysis imaging to map the northern sky with most of the large-scale structures recovered. The lunar occultation technique enables measuring the Galactic synchrotron emission integrated along the line of sight where the Moon occults the sky (e.g. \citealt{Shaver1999A&A...345..380S, McKinley2013AJ....145...23M} and McKinley et al. submitted). Future large single-dishes observing at around 150~MHz will assist further.

\section{Discussion of the derived emissivities}
\label{sec:discussion}
\begin{figure*}
\includegraphics[width=0.5\textwidth]{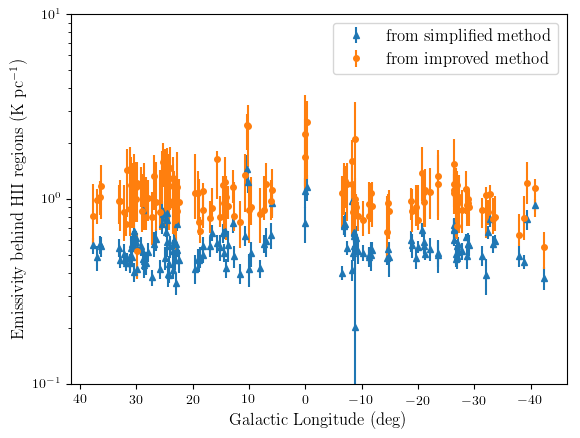}
\includegraphics[width=0.49\textwidth]{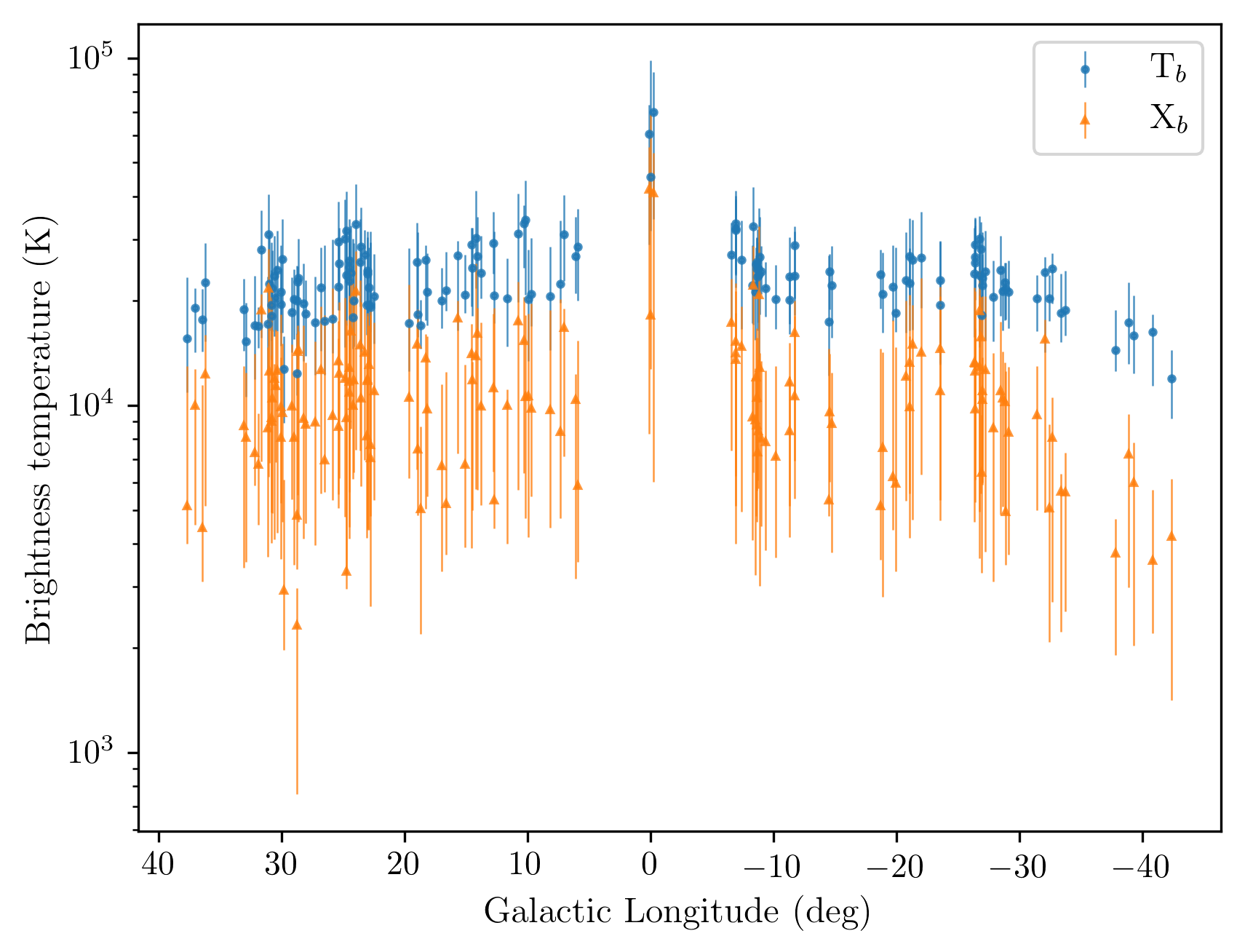}
\caption{The effect of missing short interferometric spacings on the derived emissivities at 76.2~MHz. Left: Emissivities behind \hiis from the simplified and improved methods. Right: the brightness temperature from \hiis to the Galactic edge and the brightness temperature of its missing term.}
\label{fig:comp_com_sim_glon}
\end{figure*}

\begin{figure}
\includegraphics[width=0.48\textwidth]{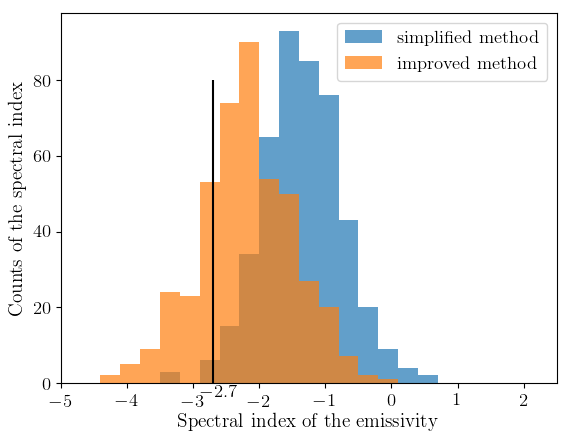}
\caption{The spectral index distribution of the derived emissivity from the simplified method and improved method. The spectral index is calculated from the emissivity (behind the \hiino) at the frequency from 76.2 to 99.2 MHz, from 83.8 to 106.9 MHz, and from 91.5 to 114.6 MHz. The bin width is 0.3.  Most of the spectral indices from the simplified method are far away  from the expected value of -2.7 shown by the black vertical line, indicating the missing flux density is affecting the simplified method. However, the emissivity from the improved method gives a spectral index close to -2.7. Note that emissivity is defined by the brightness temperature divided by a distance. For each \hiisno, its distance is a constant, so the emissivity behind that \hii and the corresponding brightness temperature should follow the same spectral index of -2.7.}
\label{fig:si_sim}
\end{figure}

\begin{figure*}
\includegraphics[width=\textwidth]{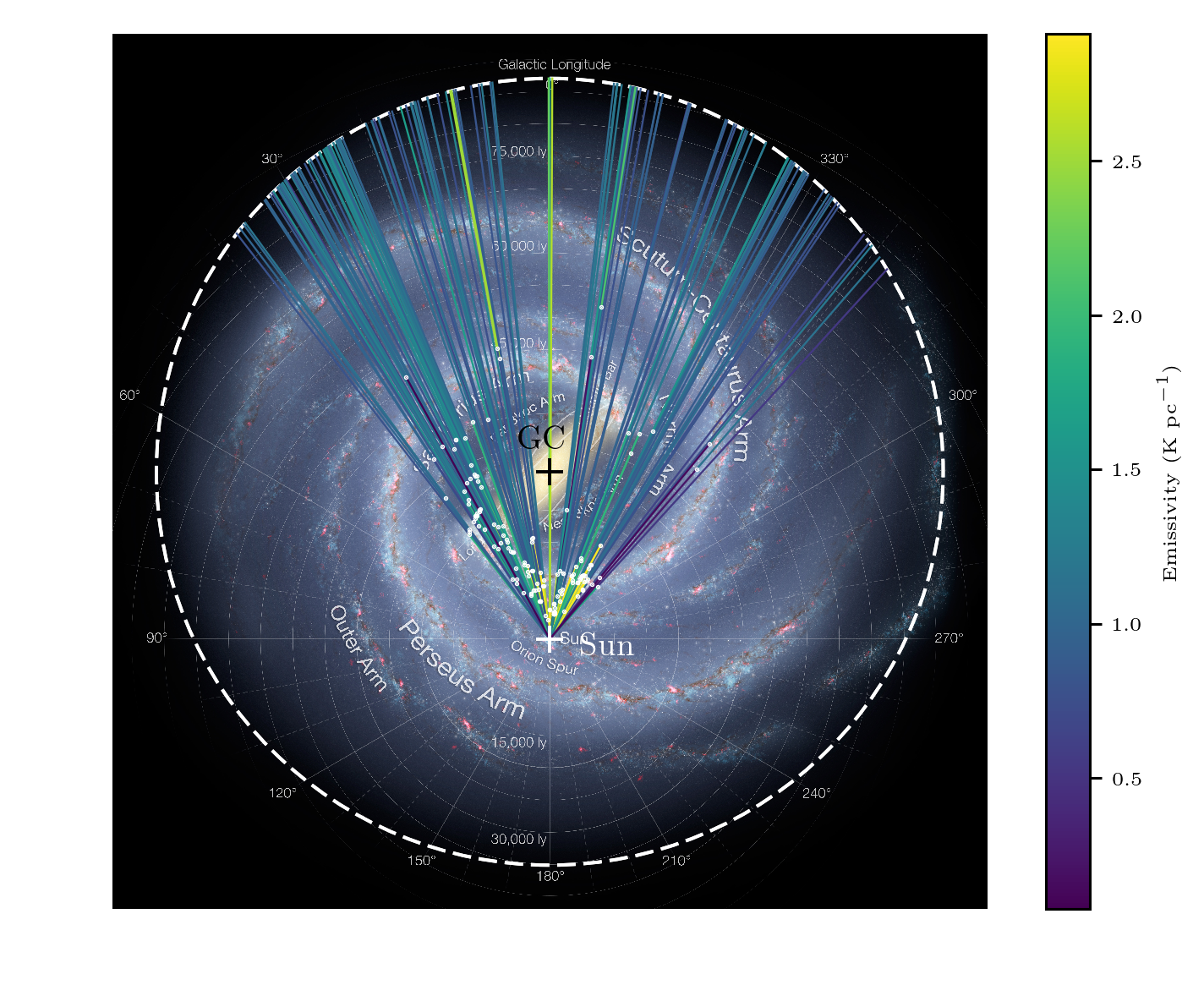}
\caption{Our new derived emissivities at 76.2~MHz both behind and in front of \hiisno. Each line indicates a path over which the emissivity is averaged with a white dot on it indicating the location of the \hiino. The background image is an artist's concept with the up-to-date information about the structures of the Milky Way. We adjusted its colour to avoid obscuring the colour of emissivities. Background image credit: NASA / JPL-Caltech / R. Hurt (SSC-Caltech) with this link: https://www.nasa.gov/jpl/charting-the-milky-way-from-the-inside-out.}
\label{fig:emi76_on_mw_img}
\end{figure*}

\begin{figure*}
\includegraphics[width=0.48\textwidth]{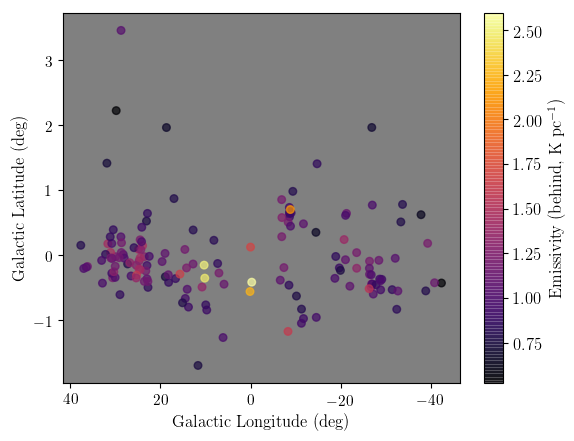}
\includegraphics[width=0.48\textwidth]{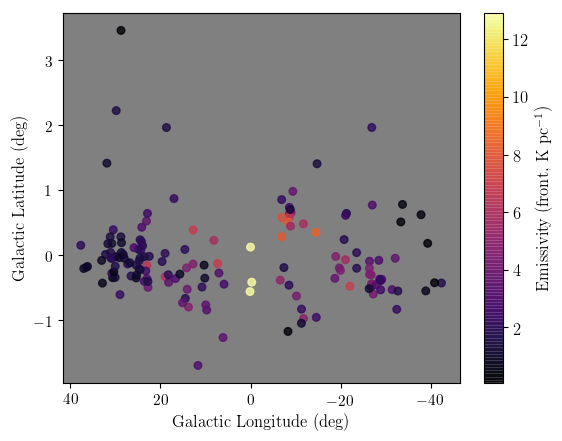}
\caption{Correctly-calculated emissivities derived from our new method from \hiis to the Galactic edge (left) and from \hiis to the Sun (right) along the line of sight.}
\label{fig:emib76_glon_glat}
\end{figure*}

\begin{figure*}
\includegraphics[width=0.48\textwidth]{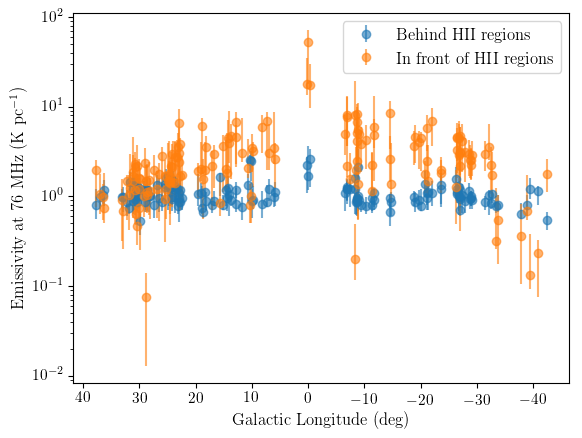}
\includegraphics[width=0.48\textwidth]{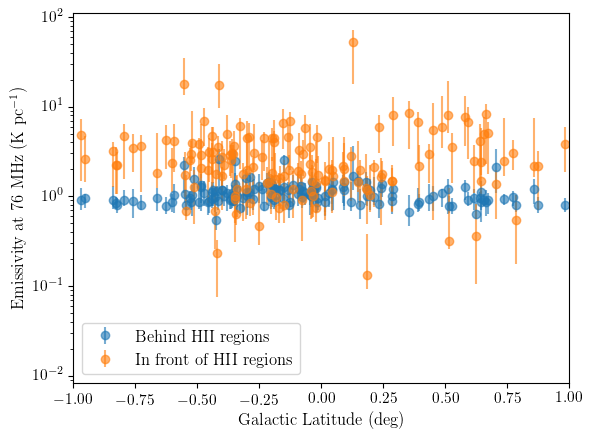}
\caption{Emissivity distribution as a function of Galactic longitude (left) and Galactic latitude (right) at 76.2~MHz. For the distribution with Galactic latitude, we only plot the emissivities derived from \hiis in the latitude range from -1 to 1 degree. }
\label{fig:emibf76}
\end{figure*}

\begin{figure}
\includegraphics[width=0.47\textwidth]{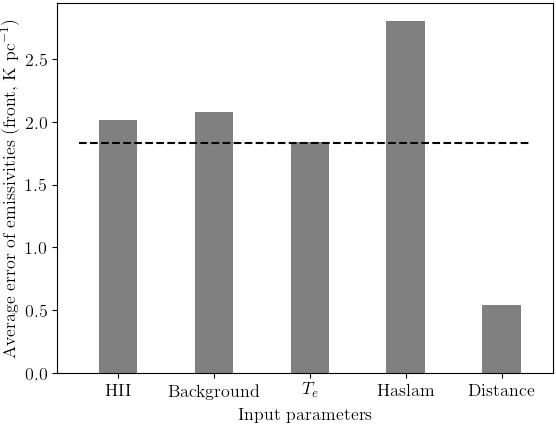}
\caption{The contribution of different input parameters to the error of the emissivities at 76.2~MHz from \hiis to the Sun. The total error of the emissivity is from different input parameters which are the rms of the absorbed region (HII), the rms of the nearby region (Background), the error of the electron temperature ($T_{\rm e}$), the rms of the background region in the Haslam map (Haslam), and the error of the distance from \hii to us (Distance). Each error here is an average of all the 152 absorption measurements. ``front'' on the y-axis means the emissivities are averaged along the path from \hii to the Sun (in front of \hiino). The error from the rms of the Haslam map contributes the most to the final error of the derived emissivities. The horizontal line indicates the average uncertainties of all the derived emissivities between the \hii and the Sun. Note that the error involved in scaling the Haslam map to our frequencies is not included here.}
\label{fig:err_from_each_input}
\end{figure}

We compare the emissivities from the simplified method and our improved method in Fig.~\ref{fig:comp_com_sim_glon} (left). The emissivities from the old method are systematically lower than those from the new method, which indicates the old method underestimates the emissivities due to the missing flux density.

We compare the total and missing brightness temperatures behind the \hii in Fig.~\ref{fig:comp_com_sim_glon} (right). 
The unrecovered brightness temperature behind \hiis ($X_{\rm b}$) is about 50\% of the total brightness temperature behind \hiis ($T_{\rm b}$), indicating that about 50\% of the large-scale structure behind \hiis has not been recovered in our observations. The brightness temperature in front of \hiis that was not recovered ($X_{\rm f}$) is comparable with the total brightness temperature in front of \hiis ($T_{\rm f}$) indicating that nearly all the large-scale structures in front of \hiis have not been recovered. Thus, the missing structures must be considered in the emissivity calculation. Note that the $X_{\rm b}$ and $X_{\rm f}$ are comparable, while $T_{\rm f}$ is about 50\% of $T_{\rm b}$. It is reasonable that most of $T_{\rm f}$ are not detected because an interferometer measures the difference along the \hii direction and its nearby direction, and also because most \hiis are nearby so that accumulated $T_{\rm f}$ is small compared to $T_{\rm b}$. The emission from the \hii to us is nearly the same for both directions, therefore, is not easily detected. However, $T_{\rm b}$ is `different' on the \hii direction and its nearby direction because most of the $T_{\rm b}$ is absorbed by the \hii on its direction, so the MWA detects a portion of $T_{\rm b}$.

\begin{figure}
\includegraphics[width=0.49\textwidth]{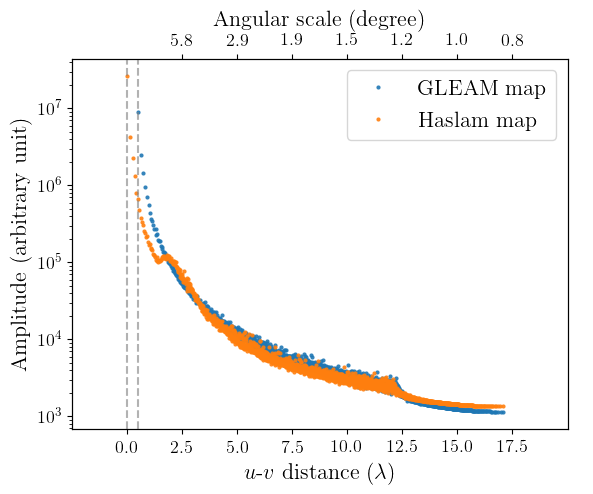}
\caption{Comparison of the visibility of the GLEAM and Haslam maps in the square region centring at $l=340^\circ$, $b=0^\circ$ with a box size of 10$^\circ$. The unit of the $u$-$v$ distance is $\lambda$ rather than k$\lambda$, because of the long wavelength of about 4 meters. The visibility data is binned (2000 bins) to show the differences clearly. The x-axis on top of the plot shows the angular size corresponding to the $u$-$v$ distance. The y-axis has an arbitrary unit, but this does not affect our comparison because they should use the same factor to make it has a physical unit. The y-axis is in log scale, so the amplitude with $u$-$v$ distance close to zero dominate the total difference of the two amplitudes. The minimum $u$-$v$ distance of the GLEAM map is small (about 0.5$\lambda$, corresponding to an angular scale of about 30$^\circ$). The $u$-$v$ distance between the two vertical lines is included in the Haslam map but is not included in the GLEAM map because of the shortest baseline of 7.7 metres. The maximum $u$-$v$ distance is the same for both maps because we smoothed them to the same resolution.  The integrated amplitude with the $u$-$v$~distance of the GLEAM map is 40\% lower than that of the Haslam map in this region. This percentage varies with regions on the Galactic plane. The average percentage of all the nine regions we checked is about 60\%.}
\label{fig:comp_vis_mwa_has_same_340_10}
\end{figure}

To confirm that the portion of missing detection is reasonable, we compare the GLEAM map and the Haslam map at the visibility plane. We use nine square regions with size of 10$^\circ$, 30$^\circ$, and 60$^\circ$ centred at $l=0^\circ$, $20^\circ$, and $340^\circ$, $b=0^\circ$.
We use the GLEAM map at the frequency of 76.2~MHz. The Haslam map is scaled from 408~MHz to the same frequency of 76.2~MHz using a spectral index of -2.7. The GLEAM map is smoothed to the same angular resolution of the Haslam map (51~arcmin), and the two maps are made with the same pixel size. For each region, we convert the two images to the visibility plane using Fast Fourier Transform (FFT) and then plot the amplitude against $u$-$v$ distance to compare the difference between the two visibilities (see Fig.~\ref{fig:comp_vis_mwa_has_same_340_10}). The difference varies with the region size and location. On average, about 60\% of the amplitude in the visibility of the Haslam map is not detected in the GLEAM survey. Our absorption analysis shows that 50\% of the large-scale structures are not recovered for the emission behind \hiisno, and nearly all emission from the column between the \hii and the Sun is not detected. These two results are generally consistent. 
 
\begin{figure}
\includegraphics[width=0.48\textwidth]{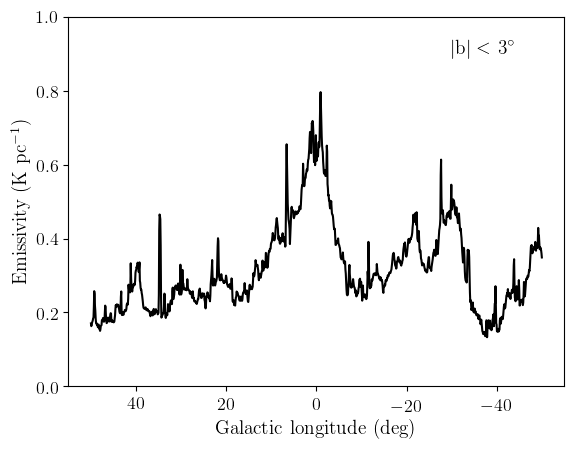}
\caption{Distribution of the measured average emissivity in the GLEAM survey along the path from the Sun to the Galactic edge with Galactic longitude from $50^\circ$ to $-50^\circ$ and latitude $|b|<3^\circ$. All detected sources and diffuse emission are included in this plot. The bin size in Galactic longitude is 4\farcm85 and in Galactic latitude is $-3^\circ < b < 3^\circ$. The Galactic centre direction has higher average emissivity compared with other directions. The existence of spiral arms possibly causes other low peaks. Note that these emissivities are directly from the GLEAM map without any correction using the Haslam map.}
\label{fig:total_emi_glon}
\end{figure}

The most apparent feature in the derived emissivities is that they increase towards the Galactic centre. Both the emissivity and the brightness temperature peak near the Galactic centre and decrease as the line-of-sight goes far away from the Galactic centre (Fig.~\ref{fig:comp_com_sim_glon}). To further confirm this trend, we check the average emissivity measured in the GLEAM map from the Sun to the Galactic edge (Fig.~\ref{fig:total_emi_glon}). It is evident that the emissivity along the path from the Sun to the Galactic edge peaks at the Galactic centre direction. 
This trend indicates the emissivity decreases with Galactocentric radius, which is modelled in \citet{Su2017MNRAS.465.3163S, Su2017MNRAS.472..828S}. This is consistent with the lowest order of disk component of the Galactic magnetic field, which is usually assumed to be exponentially distributed in the previous models (e.g. \citealt{Beuermann1985A&A...153...17B, Sun2008A&A...477..573S}). Face-on galaxies with spiral arms directly observed also show a similar profile as the one in Fig.~\ref{fig:total_emi_glon}, e.g. the LOw Frequency ARray (LOFAR; \citealt{vanHaarlem2013A&A...556A...2V}) observation of the Whirlpool Galaxy (also known as M51) at the frequency of 150~MHz (see Fig.~13 in \citealt{Mulcahy2014A&A...568A..74M}).

The average emissivities along the paths near the line of sight to the Sun are much higher than those far away from the Sun, though they have large errors. Several reasons can explain this effect. 
Firstly, the emissivity near the Galactic edge is much lower than that near the Galactic centre, which makes our average emissivities along the path high near the Galactic centre and low near the Galactic edge.  
Secondly, it may simply indicate that all distances from the \hii to the Galactic edge along the line of sight are over-estimated, which makes the emissivities behind \hiis decrease fractionally with distance. 
Thirdly, it may indicate the region near the Sun is not a representative region of the whole Milky Way because previous studies show that we are in a local bubble created by two or three supernovae \citep{Maiz-Apellaniz2001ApJ...560L..83M}, which may increase the density of cosmic-ray electrons within several kpc of the Sun.

No obvious spiral arm structures can be visually seen from our observed emissivities because the emissivity is averaged along different path lengths. Further modelling work in the future will help to reveal that whether the emissivity distribution is correlated with the spiral arms or not, because this information is embedded in our derived emissivities. From an observational aspect, we can see the spiral arms as peaks in emissivity and brightness temperature along the total paths from the Sun to the Galactic edge as a function of Galactic longitude (see Fig.~6 in \citealt{Su2017MNRAS.465.3163S} and Fig.~1 in \citealt{Beuermann1985A&A...153...17B}). 

We estimate the number density of relativistic electrons in the Galactic disk to confirm that our derived emissivities are consistent with existing electron models.
Specifically, we get the relativistic electron density by using the total power of the synchrotron emission in the Galactic disk divided by the total power of one relativistic electron and then divided by the volume of the Galactic disk.
In the above calculations, we use an average Galactic magnetic field strength of $5~\pm~1~\mu$G \citep{Sun2008A&A...477..573S} and an average emissivity of 1~$\pm$~0.5~K~pc$^{-1}$ at 76.2~MHz where 1~K~pc$^{-1}$ is  equal to 5.75$~\times~$10$^{-41}$~W~m$^{-3}$~Hz$^{-1}$~sr$^{-1}$. 
We use a typical energy of relativistic electrons of 10~$\pm$~1~GeV \citep{Stephens2001AdSpR..27..687S}, a radius of the Galactic disk of 20~kpc \citep{Nord2006AJ....132..242N}, and scale height of the Galactic disk of 1~kpc. 
We integrate the power of synchrotron emission in the frequency range 10 MHz to 1000 GHz. 
We derive a number density of relativistic electrons of 168~$\pm$~108~cm$^{-3}$. 
The relativistic electrons follow a power law distribution with energy, n$_e$(E)~=~k~E$^{-3.152}$ \citep{Adriani2017PhRvL.119r1101A}. 
Using this distribution, we derive the average density of 10~GeV electrons to be (5.6~$\pm$~3.6)~$\times$~10$^{-5}$~cm$^{-3}$, which is similar to the value of ($4~\pm~3$)~$\times~10^{-5} $ cm$^{-3}$ from the literature (see Fig.~4 in \citealt{Jansson2012ApJ...757...14J}, cited from GALPROP in \citealt{Strong2010ApJ...722L..58S}). Note that the estimated electron density has large errors due to the above typical values adopted. To further investigate the electron distribution, future work should use comprehensive Galactic magnetic field models \citep{Han2006ApJ...642..868H, Brown2007ApJ...663..258B, Sun2008A&A...477..573S, Sun2010RAA....10.1287S,  VanEck2011ApJ...728...97V}. 

\section{Summary}
\label{sec:sum}
We develop a new method of emissivity calculation by improving upon the previous simplified method. Using this new method, we calculate the synchrotron emissivities both behind and in front of 152 \hiis at six frequencies of 76.2, 83.8, 91.5, 99.2, 106.9, and 114.6~MHz. This new method enables us to derive the \hii optical depth and estimate the amount of flux density missing from our observations at each frequency. We find that the emissivities increase towards the Galactic centre. This lowest order of emissivity variation is consistent with the current Galactic magnetic field and relativistic electron distributions because both the magnetic field strength and the relativistic electron density increase towards the Galactic centre. 
The high emissivities nearby the Sun (if actually real)
might be caused by the local bubble.

The number of line-of-sight measurements will increase in the MWA phase II stage (Wayth et al., in prep.) because both the number of antenna and the maximum baselines are increased, and in the future, we will have better knowledge of  the distance and electron temperature of \hiisno. The lack of \hiis with larger distances is a key factor holding back the modelling at present because most \hiis are located near the Sun with distances less than several kpc. Future total power surveys at similar frequencies can improve the accuracy of the emissivity measurements.   
The derived emissivities may help to recover the 3-D distribution of synchrotron emission in the Milky Way. Furthermore, they provide direct information on the spatial distribution of the Galactic magnetic field and the relativistic electrons for the future modelling.

\section*{Acknowledgements}
This scientific work makes use of the Murchison Radio-astronomy Observatory, operated by CSIRO. We acknowledge the Wajarri Yamatji people as the traditional owners of the Observatory site. Support for the operation of the MWA is provided by the Australian Government (NCRIS), under a contract to Curtin University administered by Astronomy Australia Limited. We acknowledge the Pawsey Supercomputing Centre which is supported by the Western Australian and Australian Governments. HS \& WWT thanks the support from the NSFC (11473038, 11273025). This research utilized Astropy \citep{Astropy2013A&A...558A..33A}, Scipy \citep{scipy}, Numpy \citep{Walt2011arXiv1102.1523V}, and Matplotlib \citep{Hunter2007CSE.....9...90H}. We thank the anonymous referee and Denis Leahy for helpful comments.

\bibliographystyle{mnras}
\bibliography{bibtex}

\appendix

%\section{The derived emissivities}

%\newpage
\onecolumn
\begin{center}
\begin{landscape}
\renewcommand{\arraystretch}{1.5}
{\footnotesize
\begin{longtable}{@{}rrrrrrrrrrrrrrrr@{}}
\captionsetup{width=\linewidth}
\caption{The derived synchrotron emissivities and optical depths of \hiis at 76.2~MHz. An online table will show emissivities at other five frequencies of 83.8, 91.5, 99.2, 106.9, and 114.6~MHz. 
\textit{Notes.} 
Col. (1): the name of \hiis from the WISE \hii catalogue. 
Cols (2) and (3): the distance from the \hii to the Sun and the electron temperature of the \hii found in the literature \citep{Anderson2014ApJS..212....1A, Balser2015ApJ...806..199B, Hou2014A&A...569A.125H}. We use $T_{\rm e}$ = (4928 $\pm$ 277) $+$ (385 $\pm$ 29) $R_{\rm gal}$ from \citet{Balser2015ApJ...806..199B} if no electron temperature is given in the literature. 
Col. (4): the measured brightness temperature in the direction of the absorbed region.
Col. (5): the measured brightness temperature from the Sun to the Galactic edge in the absence of \hii emission (i.e. as derived from a region near the line of sight to the \hiino) from the GLEAM map.
Col. (6): the total brightness temperature (without missing flux density) from the Sun to the Galactic edge in the absence of \hii absorption derived from the Haslam map.
Col. (7): the derived brightness temperature of the synchrotron emission from \hiis to the Galactic edge. 
Col. (8): the derived brightness temperature of the synchrotron emission from the \hii to the Sun. 
Cols (9) and (10): the brightness temperature of the emission on the missing short interferometric spacings respectively between the Galactic edge and the Sun ($X_{\rm b}$), and between an \hii and the Sun ($X_{\rm f}$).
Col. (11): the optical depth of the \hiino. 
Col. (12): the average emissivity between the \hii and the Galactic edge. 
Col. (13): the average emissivity between the \hii and the Sun. 
Note that for Cols (9) and (10), the emissivities will be $\sim$15\% lower if you accept a synchrotron spectral index of -2.6 instead of -2.7 we used.
Col. (14): the emissivity between the \hii and the Galactic edge derived from the simplified method.}
 \label{tab:emi}
 \\
\hline\hline
% % % % Column names % % % %
  \multicolumn{1}{c}{\hiino} &
  \multicolumn{1}{c}{Dis} &
  \multicolumn{1}{c}{$T_{\rm e}$} &
  \multicolumn{1}{c}{$T_{\rm h}$} &
  \multicolumn{1}{c}{$T\rm _m$} &
  \multicolumn{1}{c}{$T_{\rm t}$} &
  \multicolumn{1}{c}{$T_{\rm b}$} &
  \multicolumn{1}{c}{$T_{\rm f}$} &
  \multicolumn{1}{c}{$X_{\rm b}$} &
  \multicolumn{1}{c}{$X_{\rm f}$} &
  \multicolumn{1}{c}{$\tau$} &
  \multicolumn{1}{c}{$\epsilon_{\rm b}$} &
  \multicolumn{1}{c}{$\epsilon_{\rm f}$} &
  \multicolumn{1}{c}{$\epsilon_{\rm b\_sim}$} \\
% % % % Column units % % % %
  \multicolumn{1}{c}{} &
  \multicolumn{1}{c}{kpc} &
   \multicolumn{1}{c}{$\times$10$^3$ K} &
  \multicolumn{1}{c}{$\times$10$^3$ K} & 
  \multicolumn{1}{c}{$\times$10$^3$ K} & 
  \multicolumn{1}{c}{$\times$10$^3$ K} &
  \multicolumn{1}{c}{$\times$10$^3$ K} & 
  \multicolumn{1}{c}{$\times$10$^3$ K} & 
  \multicolumn{1}{c}{$\times$10$^3$ K} & 
  \multicolumn{1}{c}{$\times$10$^3$ K} &
  \multicolumn{1}{c}{} &   
  \multicolumn{1}{c}{K pc$^{-1}$} &
  \multicolumn{1}{c}{K pc$^{-1}$} &
  \multicolumn{1}{c}{K pc$^{-1}$} \\
% % % % Column number % % % %
  \multicolumn{1}{c}{(1)} &
  \multicolumn{1}{c}{(2)} &
  \multicolumn{1}{c}{(3)} &
  \multicolumn{1}{c}{(4)} & 
  \multicolumn{1}{c}{(5)} & 
  \multicolumn{1}{c}{(6)} &
  \multicolumn{1}{c}{(7)} & 
  \multicolumn{1}{c}{(8)} & 
  \multicolumn{1}{c}{(9)} & 
  \multicolumn{1}{c}{(10)} &
  \multicolumn{1}{c}{(11)} & 
  \multicolumn{1}{c}{(12)} &  
  \multicolumn{1}{c}{(13)} &  
  \multicolumn{1}{c}{(14)} &   
  \\
\hline
\endfirsthead

\hline
% % % % Column names % % % %
  \multicolumn{1}{c}{\hiino} &
  \multicolumn{1}{c}{Dis} &
  \multicolumn{1}{c}{$T_{\rm e}$} &
  \multicolumn{1}{c}{$T_{\rm h}$} &
  \multicolumn{1}{c}{$T\rm _m$} &
  \multicolumn{1}{c}{$T_{\rm t}$} &
  \multicolumn{1}{c}{$T_{\rm b}$} &
  \multicolumn{1}{c}{$T_{\rm f}$} &
  \multicolumn{1}{c}{$X_{\rm b}$} &
  \multicolumn{1}{c}{$X_{\rm f}$} &
  \multicolumn{1}{c}{$\tau$} &
  \multicolumn{1}{c}{$\epsilon_{\rm b\_com}$} &
  \multicolumn{1}{c}{$\epsilon_{\rm f\_com}$} &
  \multicolumn{1}{c}{$\epsilon_{\rm b\_sim}$} \\
% % % % Column units % % % %
  \multicolumn{1}{c}{} &
  \multicolumn{1}{c}{kpc} &
  \multicolumn{1}{c}{$\times$10$^3$ K} &
  \multicolumn{1}{c}{$\times$10$^3$ K} & 
  \multicolumn{1}{c}{$\times$10$^3$ K} & 
  \multicolumn{1}{c}{$\times$10$^3$ K} &
  \multicolumn{1}{c}{$\times$10$^3$ K} & 
  \multicolumn{1}{c}{$\times$10$^3$ K} & 
  \multicolumn{1}{c}{$\times$10$^3$ K} & 
  \multicolumn{1}{c}{$\times$10$^3$ K} &
  \multicolumn{1}{c}{} &   
  \multicolumn{1}{c}{K pc$^{-1}$} &
  \multicolumn{1}{c}{K pc$^{-1}$} &
  \multicolumn{1}{c}{K pc$^{-1}$} \\
% % % % Column number % % % %
  \multicolumn{1}{c}{(1)} &
  \multicolumn{1}{c}{(2)} &
  \multicolumn{1}{c}{(3)} &
  \multicolumn{1}{c}{(4)} & 
  \multicolumn{1}{c}{(5)} & 
  \multicolumn{1}{c}{(6)} &
  \multicolumn{1}{c}{(7)} & 
  \multicolumn{1}{c}{(8)} & 
  \multicolumn{1}{c}{(9)} & 
  \multicolumn{1}{c}{(10)} &
  \multicolumn{1}{c}{(11)} &  
  \multicolumn{1}{c}{(12)} & 
  \multicolumn{1}{c}{(13)} &
  \multicolumn{1}{c}{(14)} &  
  \\
\hline
\endhead

 \hline
 \multicolumn{13}{r}{{Continued on next page}} \\
\endfoot
\hline \hline
\endlastfoot
G000.003$+$00.127 & 1.5~$\pm$~0.3 & 7.62~$\pm$~0.30 & 17.0~$\pm$~1.0 & 36.7~$\pm$~4.7 & 124.9~$\pm$~10.6 & 45.6$_{-13.7}^{+52.8}$ & 79.4$_{-50.3}^{+23.8}$ & 18.2$_{-5.5}^{+50.8}$ & 70.0$_{-49.7}^{+21.0}$ & 36.3 & 1.7$_{-0.5}^{+2.0}$ & 52.9$_{-35.1}^{+19.1}$ & 1.10~$\pm$~0.20 \\ 
G000.120$-$00.556 & 1.5~$\pm$~0.3 & 7.62~$\pm$~0.30 & 23.5~$\pm$~2.3 & 34.5~$\pm$~3.3 & 87.4~$\pm$~20.0 & 60.5$_{-31.5}^{+12.9}$ & 26.9$_{-4.3}^{+24.5}$ & 41.8$_{-33.5}^{+13.7}$ & 11.0$_{-3.4}^{+26.0}$ & 30.0 & 2.2$_{-1.2}^{+0.5}$ & 17.9$_{-4.6}^{+16.7}$ & 0.74~$\pm$~0.17 \\ 
G005.887$-$00.443 & 3.0~$\pm$~0.2 & 11.13~$\pm$~0.20 & 7.5~$\pm$~3.1 & 19.1~$\pm$~0.8 & 40.3~$\pm$~2.2 & 28.6$_{-8.6}^{+8.2}$ & 7.9$_{-4.2}^{+3.8}$ & 5.9$_{-2.3}^{+9.5}$ & 15.4$_{-8.8}^{+4.6}$ & 39.2 & 1.1$_{-0.3}^{+0.3}$ & 2.6$_{-1.4}^{+1.3}$ & 0.95~$\pm$~0.14 \\ 
G006.100$-$01.263 & 1.0~$\pm$~0.2 & 7.70~$\pm$~0.80 & 1.6~$\pm$~0.7 & 10.4~$\pm$~2.2 & 30.3~$\pm$~2.6 & 26.9$_{-5.9}^{+8.1}$ & 3.4$_{-0.9}^{+5.2}$ & 10.4$_{-7.2}^{+1.9}$ & 9.5$_{-2.8}^{+5.0}$ & 45.6 & 1.0$_{-0.2}^{+0.3}$ & 3.4$_{-1.1}^{+5.3}$ & 0.64~$\pm$~0.10 \\ 
G007.015$-$00.271 & 2.7~$\pm$~0.5 & 7.50~$\pm$~0.80 & 7.9~$\pm$~1.9 & 14.8~$\pm$~1.2 & 39.2~$\pm$~3.8 & 31.0$_{-8.2}^{+9.3}$ & 8.2$_{-2.5}^{+9.0}$ & 16.7$_{-9.6}^{+2.3}$ & 7.8$_{-2.2}^{+9.6}$ & 23.2 & 1.2$_{-0.3}^{+0.4}$ & 3.0$_{-1.1}^{+3.4}$ & 0.59~$\pm$~0.10 \\ 
G007.303$-$00.125 & 2.7~$\pm$~0.5 & 7.17~$\pm$~0.30 & 8.2~$\pm$~0.7 & 14.9~$\pm$~1.2 & 40.8~$\pm$~4.7 & 22.3$_{-2.6}^{+11.7}$ & 18.5$_{-12.8}^{+3.8}$ & 8.4$_{-3.7}^{+11.9}$ & 17.5$_{-11.9}^{+4.1}$ & 6.3 & 0.9$_{-0.1}^{+0.5}$ & 6.8$_{-4.9}^{+1.9}$ & 0.57~$\pm$~0.06 \\ 
G008.137$+$00.232 & 3.4~$\pm$~0.8 & 7.09~$\pm$~0.10 & 11.6~$\pm$~0.5 & 14.7~$\pm$~0.9 & 40.5~$\pm$~2.5 & 20.6$_{-6.2}^{+8.1}$ & 19.9$_{-8.5}^{+5.5}$ & 9.7$_{-5.3}^{+9.1}$ & 15.9$_{-8.6}^{+5.2}$ & $>$99 & 0.8$_{-0.3}^{+0.3}$ & 5.9$_{-2.8}^{+2.1}$ & 0.42~$\pm$~0.05 \\ 
G009.725$-$00.840 & 5.2~$\pm$~1.0 & 6.27~$\pm$~0.30 & 10.2~$\pm$~0.5 & 15.1~$\pm$~1.5 & 37.5~$\pm$~3.6 & 21.0$_{-4.1}^{+9.4}$ & 16.5$_{-9.3}^{+3.2}$ & 9.8$_{-4.3}^{+9.8}$ & 12.6$_{-9.0}^{+3.3}$ & $>$99 & 0.9$_{-0.2}^{+0.4}$ & 3.2$_{-1.9}^{+0.9}$ & 0.51~$\pm$~0.08 \\ 
G009.942$-$00.761 & 5.2~$\pm$~1.0 & 3.70~$\pm$~0.40 & 10.4~$\pm$~0.5 & 15.8~$\pm$~1.6 & 38.7~$\pm$~3.3 & 20.2$_{-6.8}^{+7.8}$ & 17.9$_{-7.5}^{+6.7}$ & 10.7$_{-6.5}^{+7.7}$ & 11.6$_{-7.3}^{+6.3}$ & 7.4 & 0.9$_{-0.3}^{+0.3}$ & 3.5$_{-1.6}^{+1.5}$ & 0.42~$\pm$~0.08 \\ 
G010.160$-$00.350 & 14.5~$\pm$~0.9 & 6.83~$\pm$~0.00 & 7.8~$\pm$~0.9 & 16.9~$\pm$~1.4 & 41.2~$\pm$~1.3 & 34.2$_{-13.4}^{+10.3}$ & 14.7$_{-7.4}^{+5.5}$ & 10.6$_{-5.8}^{+7.6}$ & 13.7$_{-7.0}^{+5.8}$ & 6.3 & 2.5$_{-1.0}^{+0.8}$ & 1.0$_{-0.5}^{+0.4}$ & 1.23~$\pm$~0.16 \\ 
G010.308$-$00.150 & 15.0~$\pm$~1.1 & 6.80~$\pm$~0.00 & 10.0~$\pm$~0.4 & 21.3~$\pm$~2.9 & 45.4~$\pm$~0.6 & 33.4$_{-8.5}^{+4.3}$ & 12.0$_{-4.3}^{+8.3}$ & 15.4$_{-9.0}^{+5.2}$ & 8.7$_{-3.7}^{+8.3}$ & 58.7 & 2.5$_{-0.7}^{+0.4}$ & 0.8$_{-0.3}^{+0.6}$ & 1.46~$\pm$~0.27 \\ 
G010.769$-$00.487 & 5.0~$\pm$~1.0 & 5.30~$\pm$~0.50 & 8.6~$\pm$~0.7 & 17.1~$\pm$~1.3 & 41.6~$\pm$~2.5 & 31.3$_{-11.5}^{+9.4}$ & 10.3$_{-1.1}^{+11.0}$ & 17.5$_{-11.8}^{+5.3}$ & 7.0$_{-2.1}^{+11.5}$ & 15.8 & 1.3$_{-0.5}^{+0.4}$ & 2.1$_{-0.5}^{+2.2}$ & 0.63~$\pm$~0.08 \\ 
G011.662$-$01.692 & 1.2~$\pm$~0.1 & 7.75~$\pm$~0.30 & 4.7~$\pm$~0.2 & 7.3~$\pm$~1.1 & 24.0~$\pm$~1.5 & 20.4$_{-5.6}^{+6.1}$ & 3.6$_{-0.7}^{+5.2}$ & 10.0$_{-6.0}^{+1.1}$ & 6.7$_{-2.0}^{+5.5}$ & 12.4 & 0.8$_{-0.2}^{+0.2}$ & 3.0$_{-0.6}^{+4.4}$ & 0.39~$\pm$~0.05 \\ 
G012.742$+$00.390 & 2.6~$\pm$~0.7 & 7.23~$\pm$~0.30 & 11.0~$\pm$~0.8 & 15.8~$\pm$~1.2 & 38.3~$\pm$~2.8 & 20.7$_{-3.5}^{+10.9}$ & 17.6$_{-10.6}^{+2.5}$ & 5.3$_{-0.9}^{+13.8}$ & 17.2$_{-13.9}^{+5.2}$ & $>$99 & 0.8$_{-0.1}^{+0.4}$ & 6.8$_{-4.5}^{+2.0}$ & 0.49~$\pm$~0.07 \\ 
G012.761$-$00.133 & 2.9~$\pm$~0.3 & 7.62~$\pm$~0.10 & 2.6~$\pm$~1.0 & 12.5~$\pm$~1.6 & 42.5~$\pm$~1.1 & 29.3$_{-5.4}^{+6.8}$ & 13.2$_{-6.9}^{+4.8}$ & 11.2$_{-4.8}^{+7.1}$ & 18.3$_{-6.3}^{+5.2}$ & 9.8 & 1.2$_{-0.2}^{+0.3}$ & 4.6$_{-2.4}^{+1.7}$ & 0.74~$\pm$~0.08 \\ 
G013.776$-$00.795 & 2.0~$\pm$~0.4 & 8.60~$\pm$~0.90 & 6.0~$\pm$~0.6 & 11.4~$\pm$~2.2 & 33.4~$\pm$~3.5 & 24.0$_{-3.6}^{+6.8}$ & 9.4$_{-6.0}^{+2.9}$ & 9.9$_{-4.8}^{+7.4}$ & 12.0$_{-6.0}^{+4.0}$ & $>$99 & 0.9$_{-0.1}^{+0.3}$ & 4.7$_{-3.1}^{+1.7}$ & 0.56~$\pm$~0.11 \\ 
G014.060$-$00.521 & 2.0~$\pm$~0.4 & 7.46~$\pm$~0.30 & 7.2~$\pm$~0.4 & 10.4~$\pm$~1.0 & 34.6~$\pm$~3.9 & 26.8$_{-11.1}^{+8.0}$ & 7.8$_{-2.6}^{+10.0}$ & 16.1$_{-10.4}^{+3.0}$ & 8.0$_{-2.9}^{+9.8}$ & $>$99 & 1.0$_{-0.4}^{+0.3}$ & 3.9$_{-1.5}^{+5.1}$ & 0.42~$\pm$~0.05 \\ 
G014.207$-$00.193 & 3.6~$\pm$~0.5 & 6.89~$\pm$~0.30 & 4.3~$\pm$~0.9 & 13.0~$\pm$~1.1 & 46.9~$\pm$~2.5 & 30.3$_{-8.6}^{+11.1}$ & 16.2$_{-11.0}^{+9.1}$ & 13.9$_{-8.1}^{+11.7}$ & 19.8$_{-10.9}^{+8.2}$ & 15.6 & 1.2$_{-0.4}^{+0.5}$ & 4.5$_{-3.1}^{+2.6}$ & 0.67~$\pm$~0.07 \\ 
G014.481$-$00.662 & 2.0~$\pm$~0.4 & 10.40~$\pm$~1.00 & 6.9~$\pm$~0.5 & 9.6~$\pm$~0.9 & 28.5~$\pm$~2.2 & 24.9$_{-6.8}^{+7.5}$ & 3.6$_{-0.9}^{+6.5}$ & 11.8$_{-6.8}^{+1.5}$ & 7.1$_{-2.1}^{+5.9}$ & 79.6 & 1.0$_{-0.3}^{+0.3}$ & 1.8$_{-0.6}^{+3.3}$ & 0.51~$\pm$~0.06 \\ 
G014.576$+$00.091 & 3.6~$\pm$~0.5 & 5.51~$\pm$~0.10 & 3.1~$\pm$~1.5 & 12.6~$\pm$~0.7 & 36.4~$\pm$~2.4 & 29.1$_{-9.8}^{+3.4}$ & 7.3$_{-2.7}^{+9.0}$ & 14.1$_{-10.2}^{+3.0}$ & 9.7$_{-2.0}^{+9.2}$ & 66.2 & 1.2$_{-0.4}^{+0.1}$ & 2.0$_{-0.8}^{+2.5}$ & 0.66~$\pm$~0.08 \\ 
G015.097$-$00.729 & 2.0~$\pm$~0.1 & 9.77~$\pm$~0.10 & 4.6~$\pm$~1.6 & 8.7~$\pm$~1.3 & 28.1~$\pm$~4.8 & 20.8$_{-2.4}^{+6.4}$ & 7.3$_{-5.0}^{+2.2}$ & 6.8$_{-2.9}^{+6.3}$ & 12.6$_{-4.9}^{+3.1}$ & $>$99 & 0.8$_{-0.1}^{+0.2}$ & 3.7$_{-2.5}^{+1.1}$ & 0.55~$\pm$~0.09 \\ 
G015.676$-$00.288 & 11.6~$\pm$~0.4 & 6.51~$\pm$~0.30 & 8.4~$\pm$~0.2 & 11.1~$\pm$~0.9 & 36.8~$\pm$~0.3 & 27.0$_{-10.6}^{+2.8}$ & 9.8$_{-2.8}^{+10.5}$ & 17.8$_{-10.6}^{+2.2}$ & 7.9$_{-2.8}^{+10.2}$ & 15.0 & 1.6$_{-0.6}^{+0.2}$ & 0.8$_{-0.2}^{+0.9}$ & 0.58~$\pm$~0.06 \\ 
G016.648$-$00.357 & 3.9~$\pm$~0.4 & 6.81~$\pm$~0.30 & 4.6~$\pm$~1.0 & 12.7~$\pm$~0.9 & 33.1~$\pm$~0.8 & 21.4$_{-2.7}^{+6.2}$ & 11.7$_{-6.7}^{+2.8}$ & 5.2$_{-1.5}^{+7.3}$ & 15.2$_{-7.7}^{+4.6}$ & $>$99 & 0.9$_{-0.1}^{+0.3}$ & 3.0$_{-1.8}^{+0.8}$ & 0.66~$\pm$~0.06 \\ 
G016.993$+$00.873 & 2.6~$\pm$~0.5 & 6.97~$\pm$~0.10 & 1.5~$\pm$~0.5 & 7.8~$\pm$~1.5 & 25.7~$\pm$~3.1 & 20.0$_{-3.4}^{+4.8}$ & 5.7$_{-3.7}^{+2.4}$ & 6.7$_{-3.4}^{+4.8}$ & 11.2$_{-3.8}^{+2.5}$ & 19.4 & 0.8$_{-0.1}^{+0.2}$ & 2.2$_{-1.5}^{+1.0}$ & 0.55~$\pm$~0.07 \\ 
G018.187$-$00.415 & 3.6~$\pm$~0.4 & 6.93~$\pm$~0.30 & 8.0~$\pm$~0.4 & 12.5~$\pm$~0.8 & 33.9~$\pm$~0.9 & 21.3$_{-4.3}^{+6.1}$ & 12.6$_{-6.2}^{+3.9}$ & 9.7$_{-4.3}^{+6.0}$ & 11.6$_{-5.9}^{+4.3}$ & 10.8 & 0.9$_{-0.2}^{+0.3}$ & 3.5$_{-1.8}^{+1.1}$ & 0.50~$\pm$~0.04 \\ 
G018.253$-$00.298 & 4.1~$\pm$~0.4 & 7.18~$\pm$~0.10 & 6.9~$\pm$~1.2 & 12.3~$\pm$~0.9 & 34.3~$\pm$~0.4 & 26.2$_{-8.7}^{+2.7}$ & 8.1$_{-3.0}^{+8.6}$ & 13.7$_{-8.6}^{+2.4}$ & 8.3$_{-2.1}^{+8.7}$ & 24.0 & 1.1$_{-0.4}^{+0.1}$ & 2.0$_{-0.8}^{+2.1}$ & 0.56~$\pm$~0.07 \\ 
G018.669$+$01.965 & 2.6~$\pm$~0.5 & 7.21~$\pm$~0.10 & 1.2~$\pm$~0.8 & 5.8~$\pm$~1.2 & 21.0~$\pm$~2.8 & 17.0$_{-2.4}^{+3.1}$ & 4.0$_{-2.4}^{+2.6}$ & 5.0$_{-2.8}^{+3.7}$ & 10.3$_{-2.5}^{+1.9}$ & 32.6 & 0.7$_{-0.1}^{+0.1}$ & 1.5$_{-1.0}^{+1.1}$ & 0.49~$\pm$~0.06 \\ 
G018.914$-$00.329 & 3.4~$\pm$~0.7 & 5.44~$\pm$~0.10 & 5.7~$\pm$~0.8 & 11.0~$\pm$~1.1 & 38.9~$\pm$~1.2 & 18.3$_{-2.6}^{+13.2}$ & 20.6$_{-13.0}^{+2.6}$ & 7.5$_{-2.7}^{+11.4}$ & 20.4$_{-11.4}^{+2.8}$ & 39.8 & 0.7$_{-0.1}^{+0.5}$ & 6.1$_{-4.0}^{+1.5}$ & 0.47~$\pm$~0.06 \\ 
G018.978$+$00.030 & 4.0~$\pm$~0.4 & 6.81~$\pm$~0.30 & 7.8~$\pm$~1.1 & 11.8~$\pm$~0.7 & 33.9~$\pm$~0.4 & 25.8$_{-7.9}^{+7.8}$ & 8.1$_{-2.1}^{+8.1}$ & 15.0$_{-8.5}^{+2.0}$ & 7.1$_{-2.5}^{+8.0}$ & 18.0 & 1.1$_{-0.3}^{+0.3}$ & 2.0$_{-0.6}^{+2.0}$ & 0.47~$\pm$~0.06 \\ 
G019.629$-$00.095 & 11.7~$\pm$~0.4 & 6.48~$\pm$~0.10 & 10.9~$\pm$~0.1 & 11.1~$\pm$~1.0 & 39.5~$\pm$~1.0 & 17.3$_{-4.7}^{+11.1}$ & 22.3$_{-11.0}^{+4.9}$ & 10.5$_{-4.4}^{+11.7}$ & 17.9$_{-11.0}^{+4.9}$ & 78.6 & 1.1$_{-0.3}^{+0.7}$ & 1.9$_{-0.9}^{+0.4}$ & 0.42~$\pm$~0.07 \\ 
G022.478$-$00.015 & 6.2~$\pm$~1.2 & 6.33~$\pm$~0.30 & 8.6~$\pm$~0.7 & 11.9~$\pm$~0.9 & 31.9~$\pm$~1.8 & 20.7$_{-5.6}^{+6.6}$ & 10.8$_{-5.6}^{+6.3}$ & 11.0$_{-5.7}^{+6.3}$ & 9.4$_{-6.6}^{+5.2}$ & 75.5 & 1.0$_{-0.3}^{+0.3}$ & 1.7$_{-1.0}^{+1.1}$ & 0.47~$\pm$~0.07 \\ 
G022.761$-$00.492 & 4.8~$\pm$~0.4 & 6.65~$\pm$~0.30 & 6.8~$\pm$~0.3 & 11.6~$\pm$~0.7 & 37.9~$\pm$~1.7 & 19.5$_{-3.2}^{+12.1}$ & 18.4$_{-12.4}^{+2.8}$ & 7.1$_{-2.3}^{+12.7}$ & 19.2$_{-13.1}^{+2.2}$ & 62.6 & 0.9$_{-0.1}^{+0.5}$ & 3.8$_{-2.6}^{+0.7}$ & 0.53~$\pm$~0.04 \\ 
G022.780$-$00.401 & 11.1~$\pm$~0.4 & 6.71~$\pm$~0.30 & 6.8~$\pm$~0.3 & 11.6~$\pm$~0.7 & 37.9~$\pm$~1.7 & 19.2$_{-4.7}^{+10.2}$ & 18.7$_{-10.9}^{+4.7}$ & 7.7$_{-5.1}^{+10.5}$ & 18.6$_{-10.6}^{+4.5}$ & 7.0 & 1.2$_{-0.3}^{+0.6}$ & 1.7$_{-1.0}^{+0.4}$ & 0.73~$\pm$~0.06 \\ 
G022.879$+$00.645 & 2.5~$\pm$~0.5 & 7.34~$\pm$~0.30 & 9.8~$\pm$~0.3 & 11.1~$\pm$~0.9 & 28.0~$\pm$~2.0 & 21.9$_{-8.8}^{+6.6}$ & 6.1$_{-1.2}^{+8.3}$ & 13.1$_{-8.7}^{+3.9}$ & 3.8$_{-1.3}^{+8.2}$ & $>$99 & 0.9$_{-0.4}^{+0.3}$ & 2.4$_{-0.7}^{+3.3}$ & 0.35~$\pm$~0.05 \\ 
G022.987$-$00.155 & 2.5~$\pm$~0.5 & 5.10~$\pm$~0.50 & 9.9~$\pm$~0.3 & 17.0~$\pm$~1.0 & 40.2~$\pm$~2.1 & 24.0$_{-6.6}^{+7.7}$ & 16.2$_{-7.0}^{+6.3}$ & 11.8$_{-7.4}^{+6.3}$ & 11.4$_{-6.1}^{+6.2}$ & 44.2 & 1.0$_{-0.3}^{+0.3}$ & 6.5$_{-3.1}^{+2.8}$ & 0.52~$\pm$~0.05 \\ 
G022.988$-$00.360 & 4.8~$\pm$~0.4 & 6.66~$\pm$~0.30 & 6.1~$\pm$~0.8 & 11.7~$\pm$~0.8 & 39.1~$\pm$~1.8 & 24.4$_{-7.6}^{+6.0}$ & 14.4$_{-5.5}^{+6.6}$ & 11.9$_{-7.5}^{+6.6}$ & 15.4$_{-6.3}^{+6.6}$ & 13.0 & 1.1$_{-0.3}^{+0.3}$ & 3.0$_{-1.2}^{+1.4}$ & 0.57~$\pm$~0.06 \\ 
G023.097$+$00.527 & 2.4~$\pm$~0.5 & 7.38~$\pm$~0.30 & 7.3~$\pm$~0.4 & 11.2~$\pm$~1.1 & 27.9~$\pm$~2.0 & 19.4$_{-3.4}^{+4.7}$ & 8.4$_{-4.8}^{+3.4}$ & 8.1$_{-4.0}^{+4.7}$ & 8.5$_{-4.7}^{+4.1}$ & 11.9 & 0.8$_{-0.1}^{+0.2}$ & 3.5$_{-2.1}^{+1.6}$ & 0.47~$\pm$~0.05 \\ 
G023.240$-$00.240 & 4.8~$\pm$~0.4 & 6.66~$\pm$~0.30 & 10.7~$\pm$~0.8 & 17.0~$\pm$~1.0 & 40.2~$\pm$~2.1 & 27.2$_{-7.8}^{+4.9}$ & 13.0$_{-5.6}^{+7.1}$ & 14.2$_{-7.2}^{+4.9}$ & 9.0$_{-5.7}^{+6.9}$ & 52.6 & 1.2$_{-0.3}^{+0.2}$ & 2.7$_{-1.2}^{+1.5}$ & 0.60~$\pm$~0.07 \\ 
G023.572$-$00.020 & 5.5~$\pm$~0.4 & 6.51~$\pm$~0.30 & 6.7~$\pm$~0.4 & 9.0~$\pm$~0.7 & 36.3~$\pm$~1.1 & 28.6$_{-14.3}^{+8.6}$ & 7.7$_{-2.3}^{+14.4}$ & 10.5$_{-4.7}^{+9.4}$ & 16.8$_{-9.3}^{+4.9}$ & $>$99 & 1.3$_{-0.6}^{+0.4}$ & 1.4$_{-0.4}^{+2.6}$ & 0.41~$\pm$~0.04 \\ 
G023.581$-$00.400 & 5.4~$\pm$~0.4 & 6.53~$\pm$~0.30 & 8.1~$\pm$~0.3 & 12.6~$\pm$~1.5 & 40.3~$\pm$~1.9 & 25.9$_{-6.6}^{+8.4}$ & 14.8$_{-8.3}^{+7.2}$ & 14.9$_{-7.5}^{+8.0}$ & 13.1$_{-7.7}^{+7.3}$ & 13.8 & 1.2$_{-0.3}^{+0.4}$ & 2.7$_{-1.5}^{+1.4}$ & 0.52~$\pm$~0.08 \\ 
G023.957$+$00.149 & 4.9~$\pm$~0.4 & 6.66~$\pm$~0.30 & 7.3~$\pm$~0.3 & 12.6~$\pm$~1.5 & 40.3~$\pm$~1.9 & 33.3$_{-13.6}^{+10.0}$ & 7.0$_{-1.4}^{+13.4}$ & 21.4$_{-13.9}^{+6.4}$ & 6.3$_{-0.9}^{+13.6}$ & 60.3 & 1.5$_{-0.6}^{+0.4}$ & 1.4$_{-0.3}^{+2.7}$ & 0.56~$\pm$~0.08 \\ 
G024.139$+$00.432 & 5.8~$\pm$~0.5 & 6.46~$\pm$~0.30 & 8.0~$\pm$~0.2 & 9.8~$\pm$~0.6 & 37.3~$\pm$~1.0 & 20.0$_{-5.3}^{+9.3}$ & 17.3$_{-9.0}^{+4.8}$ & 11.8$_{-5.4}^{+9.8}$ & 15.7$_{-8.9}^{+5.0}$ & 11.0 & 0.9$_{-0.2}^{+0.4}$ & 3.0$_{-1.6}^{+0.9}$ & 0.39~$\pm$~0.04 \\ 
G024.185$+$00.211 & 9.1~$\pm$~0.7 & 6.37~$\pm$~0.30 & 8.2~$\pm$~0.1 & 9.8~$\pm$~0.6 & 37.3~$\pm$~1.0 & 18.0$_{-3.4}^{+12.2}$ & 19.3$_{-12.4}^{+3.2}$ & 10.0$_{-3.8}^{+12.5}$ & 17.6$_{-12.2}^{+3.4}$ & 11.7 & 1.0$_{-0.2}^{+0.7}$ & 2.1$_{-1.4}^{+0.4}$ & 0.45~$\pm$~0.05 \\ 
G024.347$+$00.088 & 8.5~$\pm$~0.9 & 6.31~$\pm$~0.30 & 6.4~$\pm$~0.2 & 7.4~$\pm$~0.9 & 42.3~$\pm$~0.2 & 24.0$_{-7.9}^{+8.9}$ & 18.3$_{-9.0}^{+7.7}$ & 16.4$_{-7.8}^{+9.8}$ & 18.5$_{-9.2}^{+7.7}$ & $>$99 & 1.3$_{-0.4}^{+0.5}$ & 2.2$_{-1.1}^{+0.9}$ & 0.39~$\pm$~0.06 \\ 
G024.493$-$00.219 & 9.7~$\pm$~0.5 & 6.48~$\pm$~0.30 & 2.9~$\pm$~0.4 & 10.3~$\pm$~0.8 & 39.2~$\pm$~1.0 & 25.0$_{-6.4}^{+7.1}$ & 13.8$_{-6.0}^{+7.0}$ & 10.9$_{-6.5}^{+6.2}$ & 17.9$_{-5.7}^{+6.4}$ & 64.8 & 1.4$_{-0.4}^{+0.4}$ & 1.4$_{-0.6}^{+0.7}$ & 0.84~$\pm$~0.06 \\ 
G024.498$-$00.039 & 9.2~$\pm$~0.6 & 6.40~$\pm$~0.30 & 5.8~$\pm$~0.2 & 12.6~$\pm$~1.4 & 40.2~$\pm$~1.9 & 26.1$_{-6.7}^{+8.3}$ & 13.7$_{-7.8}^{+6.7}$ & 12.9$_{-7.2}^{+7.4}$ & 14.7$_{-8.2}^{+6.4}$ & 34.2 & 1.4$_{-0.4}^{+0.5}$ & 1.5$_{-0.9}^{+0.7}$ & 0.77~$\pm$~0.09 \\ 
G024.507$+$00.239 & 8.8~$\pm$~2.8 & 6.36~$\pm$~0.10 & 5.3~$\pm$~0.1 & 9.4~$\pm$~0.8 & 37.4~$\pm$~1.3 & 22.8$_{-7.5}^{+9.0}$ & 14.5$_{-9.2}^{+7.8}$ & 11.7$_{-7.6}^{+8.9}$ & 15.7$_{-9.0}^{+7.7}$ & 20.5 & 1.2$_{-0.4}^{+0.5}$ & 1.6$_{-1.2}^{+1.0}$ & 0.59~$\pm$~0.10 \\ 
G024.724$-$00.084 & 9.1~$\pm$~0.7 & 6.40~$\pm$~0.30 & 2.8~$\pm$~0.5 & 9.9~$\pm$~1.1 & 39.4~$\pm$~1.3 & 23.7$_{-6.7}^{+8.6}$ & 16.1$_{-8.3}^{+6.3}$ & 3.3$_{-0.4}^{+14.3}$ & 26.1$_{-13.0}^{+7.8}$ & 6.0 & 1.3$_{-0.4}^{+0.5}$ & 1.8$_{-0.9}^{+0.7}$ & 0.78~$\pm$~0.08 \\ 
G024.743$-$00.210 & 5.1~$\pm$~0.4 & 6.63~$\pm$~0.30 & 3.0~$\pm$~0.3 & 10.4~$\pm$~0.8 & 39.2~$\pm$~1.0 & 31.8$_{-14.3}^{+9.6}$ & 7.3$_{-2.2}^{+13.9}$ & 9.2$_{-5.9}^{+8.0}$ & 19.5$_{-8.5}^{+5.7}$ & 29.7 & 1.4$_{-0.6}^{+0.4}$ & 1.4$_{-0.4}^{+2.7}$ & 0.67~$\pm$~0.05 \\ 
G024.844$+$00.093 & 6.3~$\pm$~0.6 & 5.86~$\pm$~0.10 & 6.3~$\pm$~0.4 & 10.1~$\pm$~1.2 & 39.5~$\pm$~1.3 & 30.1$_{-15.4}^{+9.0}$ & 9.4$_{-2.8}^{+15.2}$ & 11.9$_{-7.2}^{+8.5}$ & 17.1$_{-8.0}^{+7.2}$ & 39.4 & 1.4$_{-0.7}^{+0.4}$ & 1.5$_{-0.5}^{+2.4}$ & 0.48~$\pm$~0.07 \\ 
G025.291$-$00.303 & 11.2~$\pm$~0.4 & 6.87~$\pm$~0.30 & 4.5~$\pm$~0.5 & 10.3~$\pm$~0.8 & 39.2~$\pm$~1.0 & 25.6$_{-6.8}^{+6.9}$ & 13.4$_{-7.1}^{+6.7}$ & 12.3$_{-6.2}^{+7.5}$ & 16.0$_{-6.5}^{+6.7}$ & 7.5 & 1.6$_{-0.4}^{+0.4}$ & 1.2$_{-0.6}^{+0.6}$ & 0.83~$\pm$~0.07 \\ 
G025.382$-$00.151 & 4.0~$\pm$~0.4 & 9.28~$\pm$~0.10 & 4.1~$\pm$~0.7 & 11.0~$\pm$~1.5 & 32.3~$\pm$~2.1 & 29.6$_{-7.7}^{+8.9}$ & 6.7$_{-3.2}^{+3.4}$ & 13.4$_{-7.9}^{+4.0}$ & 12.0$_{-2.9}^{+3.6}$ & 12.0 & 1.3$_{-0.3}^{+0.4}$ & 1.7$_{-0.8}^{+0.9}$ & 0.73~$\pm$~0.08 \\ 
G025.386$-$00.347 & 11.2~$\pm$~0.4 & 6.88~$\pm$~0.30 & 4.6~$\pm$~0.4 & 11.0~$\pm$~1.5 & 32.3~$\pm$~2.1 & 22.0$_{-3.2}^{+7.5}$ & 10.3$_{-6.9}^{+2.7}$ & 8.7$_{-3.6}^{+6.9}$ & 12.6$_{-5.9}^{+2.6}$ & 17.5 & 1.4$_{-0.2}^{+0.5}$ & 0.9$_{-0.6}^{+0.2}$ & 0.87~$\pm$~0.11 \\ 
G025.867$+$00.118 & 6.5~$\pm$~0.9 & 6.12~$\pm$~0.10 & 7.2~$\pm$~0.5 & 9.5~$\pm$~0.7 & 36.1~$\pm$~1.3 & 17.8$_{-3.6}^{+12.2}$ & 18.4$_{-12.4}^{+4.1}$ & 9.4$_{-4.0}^{+12.4}$ & 17.3$_{-12.2}^{+4.3}$ & 50.1 & 0.9$_{-0.2}^{+0.6}$ & 2.8$_{-1.9}^{+0.7}$ & 0.42~$\pm$~0.05 \\ 
G026.521$-$00.317 & 9.1~$\pm$~0.6 & 6.50~$\pm$~0.30 & 7.0~$\pm$~0.2 & 11.0~$\pm$~1.2 & 33.8~$\pm$~1.7 & 17.5$_{-5.3}^{+11.4}$ & 16.3$_{-11.4}^{+4.9}$ & 7.0$_{-1.3}^{+11.2}$ & 15.8$_{-11.0}^{+4.7}$ & 15.1 & 1.0$_{-0.3}^{+0.6}$ & 1.8$_{-1.3}^{+0.5}$ & 0.61~$\pm$~0.08 \\ 
G026.797$-$00.113 & 10.8~$\pm$~0.4 & 6.85~$\pm$~0.30 & 6.9~$\pm$~0.7 & 9.2~$\pm$~1.1 & 35.2~$\pm$~1.1 & 21.8$_{-7.5}^{+6.7}$ & 13.0$_{-6.8}^{+7.4}$ & 12.6$_{-7.1}^{+6.6}$ & 12.9$_{-6.1}^{+7.6}$ & $>$99 & 1.3$_{-0.5}^{+0.4}$ & 1.2$_{-0.6}^{+0.7}$ & 0.57~$\pm$~0.09 \\ 
G027.281$-$00.132 & 5.5~$\pm$~0.5 & 6.62~$\pm$~0.30 & 7.4~$\pm$~0.2 & 8.8~$\pm$~0.6 & 28.3~$\pm$~1.7 & 17.3$_{-5.0}^{+6.2}$ & 10.7$_{-5.5}^{+4.9}$ & 8.9$_{-5.0}^{+6.4}$ & 9.8$_{-5.2}^{+5.1}$ & 67.8 & 0.8$_{-0.2}^{+0.3}$ & 1.9$_{-1.0}^{+0.9}$ & 0.38~$\pm$~0.04 \\ 
G028.022$-$00.043 & 9.0~$\pm$~0.6 & 6.57~$\pm$~0.30 & 6.4~$\pm$~0.7 & 8.8~$\pm$~0.6 & 28.3~$\pm$~1.7 & 18.4$_{-4.5}^{+4.6}$ & 10.3$_{-5.2}^{+4.2}$ & 8.8$_{-4.2}^{+4.5}$ & 10.8$_{-5.3}^{+3.8}$ & 8.5 & 1.0$_{-0.3}^{+0.3}$ & 1.1$_{-0.6}^{+0.5}$ & 0.52~$\pm$~0.06 \\ 
G028.246$+$00.013 & 7.5~$\pm$~1.5 & 6.48~$\pm$~0.30 & 7.8~$\pm$~0.3 & 9.9~$\pm$~0.9 & 30.9~$\pm$~1.6 & 19.6$_{-6.2}^{+6.0}$ & 11.3$_{-6.3}^{+6.5}$ & 9.2$_{-5.0}^{+7.8}$ & 11.9$_{-7.9}^{+4.6}$ & 91.0 & 1.0$_{-0.3}^{+0.3}$ & 1.5$_{-0.9}^{+0.9}$ & 0.45~$\pm$~0.06 \\ 
G028.638$+$00.194 & 7.5~$\pm$~1.5 & 6.50~$\pm$~0.30 & 7.7~$\pm$~0.4 & 9.9~$\pm$~0.9 & 30.9~$\pm$~1.6 & 23.2$_{-9.3}^{+7.0}$ & 7.7$_{-2.3}^{+8.9}$ & 14.5$_{-9.9}^{+1.8}$ & 6.5$_{-2.1}^{+9.0}$ & 15.8 & 1.2$_{-0.5}^{+0.4}$ & 1.0$_{-0.4}^{+1.2}$ & 0.46~$\pm$~0.07 \\ 
G028.679$+$00.044 & 7.5~$\pm$~1.6 & 6.50~$\pm$~0.30 & 8.0~$\pm$~0.1 & 9.9~$\pm$~0.9 & 30.9~$\pm$~1.6 & 22.7$_{-9.2}^{+2.5}$ & 8.3$_{-2.8}^{+8.6}$ & 14.3$_{-8.8}^{+2.7}$ & 6.7$_{-2.8}^{+9.0}$ & 14.4 & 1.2$_{-0.5}^{+0.2}$ & 1.1$_{-0.4}^{+1.2}$ & 0.44~$\pm$~0.06 \\ 
G028.746$+$03.458 & 15.2~$\pm$~0.8 & 8.30~$\pm$~0.40 & 0.1~$\pm$~0.8 & 1.9~$\pm$~0.8 & 12.5~$\pm$~0.8 & 12.4$_{-1.7}^{+3.7}$ & 1.1$_{-0.9}^{+1.0}$ & 2.3$_{-1.6}^{+0.6}$ & 8.3$_{-2.5}^{+2.1}$ & 18.1 & 1.0$_{-0.2}^{+0.3}$ & 0.1$_{-0.1}^{+0.1}$ & 0.87~$\pm$~0.13 \\ 
G028.774$+$00.285 & 7.4~$\pm$~1.5 & 6.50~$\pm$~0.30 & 7.5~$\pm$~0.2 & 10.0~$\pm$~0.9 & 30.4~$\pm$~1.7 & 20.0$_{-6.1}^{+5.2}$ & 10.8$_{-5.4}^{+5.4}$ & 4.8$_{-1.4}^{+11.2}$ & 15.5$_{-10.7}^{+4.7}$ & 14.2 & 1.0$_{-0.3}^{+0.3}$ & 1.5$_{-0.8}^{+0.8}$ & 0.48~$\pm$~0.06 \\ 
G028.983$-$00.604 & 3.5~$\pm$~0.4 & 7.12~$\pm$~0.30 & 5.3~$\pm$~0.3 & 10.0~$\pm$~0.6 & 28.2~$\pm$~1.5 & 20.3$_{-4.7}^{+4.4}$ & 8.3$_{-4.3}^{+3.8}$ & 8.1$_{-4.6}^{+4.5}$ & 10.4$_{-4.1}^{+3.6}$ & $>$99 & 0.9$_{-0.2}^{+0.2}$ & 2.4$_{-1.2}^{+1.1}$ & 0.53~$\pm$~0.04 \\ 
G029.165$-$00.035 & 11.2~$\pm$~0.4 & 7.09~$\pm$~0.30 & 8.5~$\pm$~0.2 & 10.0~$\pm$~0.9 & 30.4~$\pm$~1.7 & 18.5$_{-4.3}^{+4.8}$ & 11.2$_{-4.7}^{+5.2}$ & 9.9$_{-4.6}^{+5.0}$ & 9.9$_{-4.6}^{+5.1}$ & 16.7 & 1.2$_{-0.3}^{+0.3}$ & 1.0$_{-0.4}^{+0.5}$ & 0.56~$\pm$~0.07 \\ 
G029.816$+$02.225 & 2.7~$\pm$~0.4 & 7.35~$\pm$~0.30 & 2.8~$\pm$~0.2 & 5.3~$\pm$~0.8 & 17.0~$\pm$~1.1 & 12.7$_{-3.8}^{+3.1}$ & 4.2$_{-3.5}^{+1.3}$ & 2.9$_{-1.0}^{+3.2}$ & 8.8$_{-3.1}^{+2.6}$ & 29.9 & 0.5$_{-0.2}^{+0.1}$ & 1.6$_{-1.3}^{+0.5}$ & 0.42~$\pm$~0.04 \\ 
G029.956$-$00.020 & 5.3~$\pm$~0.5 & 6.75~$\pm$~0.30 & 5.6~$\pm$~0.2 & 10.0~$\pm$~0.9 & 30.4~$\pm$~1.7 & 26.4$_{-10.2}^{+7.9}$ & 4.0$_{-1.2}^{+9.9}$ & 9.5$_{-4.9}^{+5.5}$ & 11.0$_{-5.4}^{+4.2}$ & 23.2 & 1.2$_{-0.5}^{+0.4}$ & 0.7$_{-0.2}^{+1.9}$ & 0.54~$\pm$~0.05 \\ 
G030.036$-$00.167 & 8.7~$\pm$~0.8 & 6.65~$\pm$~0.30 & 5.7~$\pm$~0.1 & 9.8~$\pm$~0.6 & 35.3~$\pm$~0.3 & 21.3$_{-5.1}^{+7.6}$ & 14.0$_{-7.7}^{+4.9}$ & 9.9$_{-4.6}^{+8.4}$ & 15.6$_{-8.5}^{+4.6}$ & 17.4 & 1.2$_{-0.3}^{+0.4}$ & 1.6$_{-0.9}^{+0.6}$ & 0.62~$\pm$~0.05 \\ 
G030.055$-$00.339 & 7.4~$\pm$~1.5 & 6.57~$\pm$~0.30 & 5.5~$\pm$~0.1 & 9.9~$\pm$~0.7 & 28.3~$\pm$~2.3 & 19.5$_{-4.7}^{+5.2}$ & 8.9$_{-4.2}^{+3.5}$ & 8.1$_{-4.5}^{+5.7}$ & 10.7$_{-4.7}^{+3.3}$ & 23.6 & 1.0$_{-0.3}^{+0.3}$ & 1.2$_{-0.6}^{+0.5}$ & 0.59~$\pm$~0.06 \\ 
G030.338$-$00.252 & 8.2~$\pm$~2.8 & 6.61~$\pm$~0.30 & 4.6~$\pm$~0.2 & 9.9~$\pm$~0.7 & 28.3~$\pm$~2.3 & 24.5$_{-8.1}^{+7.3}$ & 3.8$_{-0.6}^{+8.1}$ & 12.6$_{-8.3}^{+3.8}$ & 5.8$_{-1.7}^{+8.3}$ & 19.5 & 1.3$_{-0.5}^{+0.4}$ & 0.5$_{-0.2}^{+1.0}$ & 0.67~$\pm$~0.11 \\ 
G030.468$+$00.394 & 3.3~$\pm$~0.7 & 7.20~$\pm$~0.30 & 7.2~$\pm$~0.3 & 9.3~$\pm$~0.9 & 27.9~$\pm$~2.4 & 20.4$_{-4.0}^{+4.8}$ & 7.3$_{-4.0}^{+4.7}$ & 11.4$_{-4.5}^{+5.1}$ & 7.1$_{-3.5}^{+4.8}$ & 14.7 & 0.9$_{-0.2}^{+0.2}$ & 2.2$_{-1.3}^{+1.5}$ & 0.40~$\pm$~0.05 \\ 
G030.599$-$00.344 & 7.3~$\pm$~0.1 & 6.59~$\pm$~0.30 & 4.8~$\pm$~0.1 & 9.9~$\pm$~0.7 & 28.3~$\pm$~2.3 & 23.6$_{-7.8}^{+7.1}$ & 4.7$_{-1.2}^{+7.7}$ & 11.9$_{-7.8}^{+1.2}$ & 6.5$_{-0.8}^{+7.8}$ & 67.1 & 1.2$_{-0.4}^{+0.4}$ & 0.6$_{-0.2}^{+1.1}$ & 0.63~$\pm$~0.05 \\ 
G030.758$-$00.047 & 6.5~$\pm$~0.7 & 7.03~$\pm$~0.00 & 4.5~$\pm$~0.2 & 8.6~$\pm$~0.5 & 36.4~$\pm$~0.6 & 21.8$_{-5.4}^{+6.2}$ & 14.7$_{-6.4}^{+5.1}$ & 10.5$_{-5.6}^{+6.0}$ & 17.6$_{-6.2}^{+5.0}$ & 13.3 & 1.1$_{-0.3}^{+0.3}$ & 2.3$_{-1.0}^{+0.8}$ & 0.57~$\pm$~0.04 \\ 
G030.795$-$00.275 & 7.3~$\pm$~0.5 & 6.60~$\pm$~0.30 & 5.6~$\pm$~0.2 & 8.1~$\pm$~0.7 & 33.5~$\pm$~3.5 & 18.1$_{-4.6}^{+11.3}$ & 15.3$_{-10.8}^{+4.3}$ & 9.0$_{-4.1}^{+11.6}$ & 16.4$_{-10.1}^{+4.3}$ & 15.8 & 0.9$_{-0.2}^{+0.6}$ & 2.1$_{-1.5}^{+0.6}$ & 0.48~$\pm$~0.05 \\ 
G030.796$+$00.183 & 7.3~$\pm$~0.7 & 6.60~$\pm$~0.30 & 5.6~$\pm$~0.1 & 9.3~$\pm$~0.9 & 27.9~$\pm$~2.4 & 19.4$_{-4.0}^{+5.1}$ & 8.6$_{-4.9}^{+3.6}$ & 9.2$_{-5.2}^{+4.8}$ & 9.6$_{-4.5}^{+4.1}$ & 67.5 & 1.0$_{-0.2}^{+0.3}$ & 1.2$_{-0.7}^{+0.5}$ & 0.55~$\pm$~0.06 \\ 
G031.016$-$00.039 & 8.0~$\pm$~2.8 & 6.64~$\pm$~0.30 & 5.8~$\pm$~0.1 & 9.0~$\pm$~0.4 & 37.4~$\pm$~0.9 & 22.4$_{-5.9}^{+8.8}$ & 15.0$_{-9.0}^{+5.9}$ & 12.5$_{-5.7}^{+8.8}$ & 15.9$_{-8.4}^{+6.0}$ & 11.9 & 1.2$_{-0.4}^{+0.5}$ & 1.9$_{-1.3}^{+1.0}$ & 0.54~$\pm$~0.09 \\ 
G031.065$+$00.045 & 5.5~$\pm$~1.1 & 6.75~$\pm$~0.30 & 5.7~$\pm$~0.1 & 8.4~$\pm$~0.7 & 40.4~$\pm$~2.0 & 31.1$_{-16.2}^{+9.3}$ & 9.3$_{-2.8}^{+15.7}$ & 21.7$_{-15.5}^{+6.5}$ & 10.3$_{-3.1}^{+15.8}$ & $>$99 & 1.5$_{-0.8}^{+0.4}$ & 1.7$_{-0.6}^{+2.9}$ & 0.46~$\pm$~0.05 \\ 
G031.138$+$00.285 & 7.3~$\pm$~0.1 & 6.62~$\pm$~0.30 & 7.4~$\pm$~0.2 & 9.3~$\pm$~0.9 & 27.9~$\pm$~2.4 & 17.2$_{-4.8}^{+6.3}$ & 10.7$_{-5.8}^{+4.2}$ & 8.6$_{-5.0}^{+6.1}$ & 10.0$_{-5.8}^{+4.3}$ & 17.8 & 0.9$_{-0.2}^{+0.3}$ & 1.5$_{-0.8}^{+0.6}$ & 0.45~$\pm$~0.05 \\ 
G031.677$+$00.179 & 7.2~$\pm$~0.1 & 6.65~$\pm$~0.30 & 6.5~$\pm$~0.2 & 8.9~$\pm$~0.4 & 36.9~$\pm$~0.9 & 28.0$_{-12.4}^{+8.4}$ & 8.9$_{-1.8}^{+12.6}$ & 18.8$_{-12.0}^{+2.1}$ & 9.2$_{-1.8}^{+12.3}$ & $>$99 & 1.4$_{-0.6}^{+0.4}$ & 1.2$_{-0.2}^{+1.7}$ & 0.48~$\pm$~0.03 \\ 
G031.881$+$01.417 & 3.6~$\pm$~0.4 & 7.15~$\pm$~0.30 & 5.0~$\pm$~0.3 & 7.9~$\pm$~0.8 & 20.6~$\pm$~1.2 & 16.9$_{-2.2}^{+2.3}$ & 3.7$_{-2.1}^{+2.5}$ & 6.7$_{-2.2}^{+2.7}$ & 6.0$_{-2.2}^{+2.7}$ & 16.3 & 0.7$_{-0.1}^{+0.1}$ & 1.0$_{-0.6}^{+0.7}$ & 0.45~$\pm$~0.04 \\ 
G032.181$+$00.015 & 6.8~$\pm$~0.7 & 6.68~$\pm$~0.30 & 6.6~$\pm$~0.1 & 9.6~$\pm$~1.0 & 26.9~$\pm$~1.8 & 17.0$_{-5.1}^{+6.6}$ & 10.0$_{-6.2}^{+1.5}$ & 7.3$_{-1.4}^{+6.7}$ & 10.0$_{-6.0}^{+1.9}$ & 20.8 & 0.9$_{-0.3}^{+0.3}$ & 1.5$_{-0.9}^{+0.3}$ & 0.50~$\pm$~0.06 \\ 
G032.870$-$00.427 & 10.9~$\pm$~0.4 & 6.07~$\pm$~0.20 & 5.5~$\pm$~0.0 & 6.7~$\pm$~0.6 & 22.7~$\pm$~1.5 & 15.3$_{-4.7}^{+4.5}$ & 7.4$_{-4.5}^{+3.9}$ & 8.1$_{-4.5}^{+4.3}$ & 7.9$_{-4.4}^{+4.1}$ & 27.8 & 1.0$_{-0.3}^{+0.3}$ & 0.7$_{-0.4}^{+0.4}$ & 0.47~$\pm$~0.05 \\ 
G033.051$-$00.078 & 7.1~$\pm$~1.0 & 6.71~$\pm$~0.30 & 5.6~$\pm$~0.3 & 9.1~$\pm$~0.8 & 25.4~$\pm$~1.4 & 18.9$_{-4.6}^{+4.3}$ & 6.4$_{-4.1}^{+4.7}$ & 8.7$_{-5.3}^{+4.2}$ & 7.5$_{-3.9}^{+4.7}$ & 17.0 & 1.0$_{-0.2}^{+0.2}$ & 0.9$_{-0.6}^{+0.7}$ & 0.55~$\pm$~0.06 \\ 
G036.192$-$00.171 & 6.9~$\pm$~1.7 & 6.86~$\pm$~0.30 & 7.7~$\pm$~0.2 & 11.1~$\pm$~0.8 & 27.7~$\pm$~0.4 & 22.6$_{-7.3}^{+6.8}$ & 5.1$_{-1.0}^{+7.3}$ & 12.3$_{-7.2}^{+3.7}$ & 4.3$_{-0.6}^{+7.3}$ & 25.1 & 1.2$_{-0.4}^{+0.4}$ & 0.7$_{-0.2}^{+1.1}$ & 0.55~$\pm$~0.07 \\ 
G036.459$-$00.183 & 8.9~$\pm$~0.6 & 7.03~$\pm$~0.30 & 7.8~$\pm$~0.2 & 10.3~$\pm$~0.9 & 26.6~$\pm$~0.2 & 17.7$_{-3.4}^{+3.9}$ & 8.8$_{-3.9}^{+3.4}$ & 4.5$_{-1.3}^{+7.0}$ & 11.8$_{-7.0}^{+3.5}$ & 2.4 & 1.0$_{-0.2}^{+0.2}$ & 1.0$_{-0.4}^{+0.4}$ & 0.57~$\pm$~0.06 \\ 
G037.028$-$00.202 & 6.8~$\pm$~1.2 & 6.90~$\pm$~0.30 & 7.5~$\pm$~0.2 & 9.7~$\pm$~1.1 & 26.4~$\pm$~0.3 & 19.1$_{-4.8}^{+2.7}$ & 7.3$_{-2.7}^{+4.7}$ & 10.0$_{-5.4}^{+2.7}$ & 6.6$_{-2.4}^{+5.0}$ & 29.0 & 1.0$_{-0.3}^{+0.2}$ & 1.1$_{-0.4}^{+0.7}$ & 0.48~$\pm$~0.08 \\ 
G037.677$+$00.155 & 6.7~$\pm$~0.1 & 6.93~$\pm$~0.30 & 6.5~$\pm$~0.5 & 10.0~$\pm$~0.8 & 28.5~$\pm$~0.8 & 15.6$_{-4.7}^{+7.8}$ & 13.0$_{-7.9}^{+3.9}$ & 5.1$_{-1.1}^{+7.8}$ & 13.4$_{-7.9}^{+4.0}$ & 52.5 & 0.8$_{-0.2}^{+0.4}$ & 1.9$_{-1.2}^{+0.6}$ & 0.56~$\pm$~0.06 \\ 
G317.628$-$00.425 & 3.6~$\pm$~0.7 & 4.50~$\pm$~0.50 & 6.4~$\pm$~0.3 & 9.7~$\pm$~0.8 & 18.3~$\pm$~1.6 & 12.0$_{-2.8}^{+2.5}$ & 6.3$_{-1.9}^{+2.9}$ & 4.2$_{-2.8}^{+2.0}$ & 4.4$_{-2.0}^{+3.0}$ & 41.8 & 0.5$_{-0.1}^{+0.1}$ & 1.8$_{-0.6}^{+0.9}$ & 0.37~$\pm$~0.05 \\ 
G319.164$-$00.421 & 11.4~$\pm$~0.6 & 7.80~$\pm$~0.40 & 4.1~$\pm$~0.5 & 9.0~$\pm$~0.5 & 18.9~$\pm$~1.0 & 16.3$_{-4.9}^{+2.0}$ & 2.7$_{-1.8}^{+1.1}$ & 3.6$_{-1.4}^{+2.1}$ & 6.3$_{-1.7}^{+1.3}$ & 12.4 & 1.1$_{-0.3}^{+0.1}$ & 0.2$_{-0.2}^{+0.1}$ & 0.93~$\pm$~0.07 \\ 
G320.692$+$00.185 & 12.8~$\pm$~0.7 & 8.10~$\pm$~0.40 & 5.4~$\pm$~0.2 & 7.2~$\pm$~1.0 & 17.6~$\pm$~0.8 & 15.9$_{-3.6}^{+4.8}$ & 1.7$_{-0.5}^{+3.2}$ & 6.0$_{-4.0}^{+1.8}$ & 4.4$_{-1.3}^{+3.3}$ & $>$99 & 1.2$_{-0.3}^{+0.4}$ & 0.1$_{-0.0}^{+0.2}$ & 0.78~$\pm$~0.10 \\ 
G321.115$-$00.546 & 4.0~$\pm$~0.6 & 7.22~$\pm$~0.30 & 5.2~$\pm$~0.3 & 7.7~$\pm$~0.6 & 20.1~$\pm$~0.6 & 17.4$_{-4.2}^{+5.2}$ & 2.8$_{-0.4}^{+4.2}$ & 7.2$_{-4.2}^{+2.2}$ & 5.1$_{-1.5}^{+4.1}$ & 32.1 & 0.8$_{-0.2}^{+0.2}$ & 0.7$_{-0.1}^{+1.0}$ & 0.46~$\pm$~0.04 \\ 
G322.162$+$00.625 & 3.5~$\pm$~0.5 & 7.29~$\pm$~0.30 & 3.4~$\pm$~0.5 & 6.8~$\pm$~1.2 & 15.7~$\pm$~0.8 & 14.5$_{-1.9}^{+4.3}$ & 1.3$_{-0.9}^{+1.6}$ & 3.7$_{-1.8}^{+1.0}$ & 5.2$_{-0.7}^{+1.7}$ & 19.8 & 0.6$_{-0.1}^{+0.2}$ & 0.4$_{-0.3}^{+0.5}$ & 0.49~$\pm$~0.07 \\ 
G326.270$+$00.783 & 3.0~$\pm$~0.4 & 7.33~$\pm$~0.30 & 3.3~$\pm$~0.6 & 9.2~$\pm$~0.6 & 19.1~$\pm$~0.9 & 18.8$_{-2.9}^{+5.6}$ & 1.6$_{-1.1}^{+1.5}$ & 5.6$_{-3.1}^{+1.7}$ & 5.7$_{-1.4}^{+1.4}$ & 9.2 & 0.8$_{-0.1}^{+0.2}$ & 0.5$_{-0.4}^{+0.5}$ & 0.59~$\pm$~0.04 \\ 
G326.643$+$00.514 & 3.0~$\pm$~0.4 & 7.32~$\pm$~0.30 & 3.8~$\pm$~1.0 & 9.3~$\pm$~0.7 & 19.4~$\pm$~1.1 & 18.4$_{-3.2}^{+5.5}$ & 0.9$_{-0.1}^{+2.8}$ & 5.6$_{-3.4}^{+0.7}$ & 4.4$_{-1.3}^{+2.9}$ & $>$99 & 0.8$_{-0.1}^{+0.2}$ & 0.3$_{-0.1}^{+0.9}$ & 0.57~$\pm$~0.06 \\ 
G327.300$-$00.548 & 3.5~$\pm$~0.7 & 4.70~$\pm$~0.50 & 4.5~$\pm$~0.8 & 16.5~$\pm$~1.1 & 30.7~$\pm$~1.8 & 24.7$_{-5.1}^{+2.6}$ & 6.0$_{-2.3}^{+4.8}$ & 8.1$_{-5.4}^{+2.5}$ & 6.2$_{-2.1}^{+4.6}$ & 5.8 & 1.1$_{-0.2}^{+0.1}$ & 1.7$_{-0.7}^{+1.4}$ & 0.78~$\pm$~0.07 \\ 
G327.555$-$00.829 & 2.6~$\pm$~0.5 & 7.41~$\pm$~0.30 & 5.4~$\pm$~0.3 & 13.0~$\pm$~1.0 & 26.6~$\pm$~1.8 & 20.3$_{-2.8}^{+3.4}$ & 5.8$_{-2.8}^{+3.2}$ & 5.1$_{-3.0}^{+3.8}$ & 8.5$_{-3.5}^{+2.9}$ & 17.8 & 0.8$_{-0.1}^{+0.1}$ & 2.2$_{-1.2}^{+1.3}$ & 0.66~$\pm$~0.05 \\ 
G327.889$-$00.045 & 3.6~$\pm$~0.7 & 6.00~$\pm$~0.60 & 11.7~$\pm$~0.3 & 14.3~$\pm$~1.6 & 37.0~$\pm$~0.2 & 24.1$_{-9.9}^{+2.6}$ & 12.9$_{-2.6}^{+9.9}$ & 15.5$_{-10.6}^{+2.2}$ & 7.2$_{-2.1}^{+10.3}$ & $>$99 & 1.0$_{-0.4}^{+0.1}$ & 3.6$_{-1.0}^{+2.8}$ & 0.39~$\pm$~0.09 \\ 
G328.572$-$00.527 & 3.4~$\pm$~0.4 & 7.19~$\pm$~0.30 & 12.6~$\pm$~0.5 & 16.5~$\pm$~1.1 & 30.7~$\pm$~1.8 & 20.3$_{-4.0}^{+3.5}$ & 10.1$_{-3.4}^{+3.8}$ & 9.4$_{-4.4}^{+3.6}$ & 5.0$_{-3.4}^{+3.9}$ & 41.1 & 0.9$_{-0.2}^{+0.1}$ & 3.0$_{-1.1}^{+1.2}$ & 0.49~$\pm$~0.06 \\ 
G330.873$-$00.369 & 3.7~$\pm$~0.4 & 7.07~$\pm$~0.30 & 10.7~$\pm$~0.3 & 16.1~$\pm$~1.5 & 32.2~$\pm$~1.5 & 21.2$_{-4.5}^{+4.9}$ & 10.7$_{-4.3}^{+4.5}$ & 8.4$_{-4.7}^{+4.5}$ & 7.7$_{-4.2}^{+4.1}$ & 75.2 & 0.9$_{-0.2}^{+0.2}$ & 2.9$_{-1.2}^{+1.3}$ & 0.56~$\pm$~0.07 \\ 
G331.123$-$00.530 & 4.3~$\pm$~0.4 & 6.92~$\pm$~0.30 & 10.7~$\pm$~0.4 & 16.1~$\pm$~1.5 & 32.2~$\pm$~1.5 & 21.6$_{-4.0}^{+4.1}$ & 11.0$_{-4.4}^{+3.6}$ & 5.0$_{-1.5}^{+7.6}$ & 7.5$_{-4.1}^{+3.4}$ & $>$99 & 1.0$_{-0.2}^{+0.2}$ & 2.6$_{-1.1}^{+0.9}$ & 0.57~$\pm$~0.08 \\ 
G331.156$-$00.391 & 4.3~$\pm$~0.4 & 6.92~$\pm$~0.30 & 10.6~$\pm$~0.5 & 16.1~$\pm$~1.5 & 32.2~$\pm$~1.5 & 22.6$_{-5.0}^{+3.2}$ & 9.6$_{-2.8}^{+4.9}$ & 10.2$_{-5.3}^{+3.1}$ & 5.9$_{-2.6}^{+5.1}$ & 24.2 & 1.0$_{-0.2}^{+0.1}$ & 2.2$_{-0.7}^{+1.2}$ & 0.57~$\pm$~0.08 \\ 
G331.338$-$00.365 & 4.1~$\pm$~0.4 & 6.96~$\pm$~0.30 & 12.1~$\pm$~0.4 & 15.9~$\pm$~1.4 & 31.8~$\pm$~1.5 & 21.3$_{-4.9}^{+4.0}$ & 10.5$_{-3.9}^{+4.6}$ & 10.5$_{-5.4}^{+3.8}$ & 5.4$_{-3.8}^{+4.9}$ & 17.6 & 0.9$_{-0.2}^{+0.2}$ & 2.6$_{-1.0}^{+1.1}$ & 0.49~$\pm$~0.07 \\ 
G331.520$-$00.076 & 5.4~$\pm$~0.5 & 6.68~$\pm$~0.30 & 13.6~$\pm$~0.2 & 19.6~$\pm$~0.3 & 40.4~$\pm$~0.7 & 24.5$_{-6.4}^{+6.3}$ & 15.8$_{-6.5}^{+6.7}$ & 11.0$_{-6.2}^{+6.4}$ & 10.1$_{-6.9}^{+6.0}$ & 45.2 & 1.1$_{-0.3}^{+0.3}$ & 2.9$_{-1.2}^{+1.3}$ & 0.62~$\pm$~0.03 \\ 
G332.145$-$00.452 & 3.7~$\pm$~0.4 & 7.05~$\pm$~0.30 & 9.8~$\pm$~0.5 & 14.5~$\pm$~1.0 & 32.6~$\pm$~2.3 & 20.5$_{-5.2}^{+5.6}$ & 11.5$_{-6.0}^{+5.0}$ & 8.6$_{-5.5}^{+5.6}$ & 9.3$_{-6.0}^{+4.9}$ & 9.2 & 0.9$_{-0.2}^{+0.2}$ & 3.1$_{-1.7}^{+1.4}$ & 0.52~$\pm$~0.06 \\ 
G332.762$-$00.595 & 3.6~$\pm$~0.4 & 7.06~$\pm$~0.30 & 7.5~$\pm$~0.8 & 12.1~$\pm$~0.7 & 39.1~$\pm$~3.6 & 24.3$_{-8.9}^{+7.6}$ & 14.8$_{-7.9}^{+8.4}$ & 12.7$_{-8.9}^{+8.0}$ & 14.3$_{-7.6}^{+8.2}$ & $>$99 & 1.0$_{-0.4}^{+0.3}$ & 4.1$_{-2.2}^{+2.4}$ & 0.52~$\pm$~0.05 \\ 
G332.978$+$00.773 & 3.5~$\pm$~0.4 & 7.09~$\pm$~0.30 & 11.6~$\pm$~0.6 & 17.3~$\pm$~0.9 & 33.8~$\pm$~1.6 & 23.2$_{-4.6}^{+4.0}$ & 10.5$_{-3.4}^{+4.3}$ & 10.4$_{-4.4}^{+3.3}$ & 6.0$_{-3.1}^{+4.5}$ & 22.5 & 1.0$_{-0.2}^{+0.2}$ & 3.0$_{-1.0}^{+1.3}$ & 0.57~$\pm$~0.06 \\ 
G333.011$-$00.441 & 3.6~$\pm$~0.4 & 7.06~$\pm$~0.30 & 8.0~$\pm$~0.2 & 12.1~$\pm$~0.7 & 39.0~$\pm$~3.7 & 22.1$_{-4.5}^{+9.5}$ & 16.9$_{-10.4}^{+4.0}$ & 11.0$_{-5.1}^{+9.5}$ & 15.9$_{-10.5}^{+4.1}$ & 18.8 & 0.9$_{-0.2}^{+0.4}$ & 4.7$_{-2.9}^{+1.2}$ & 0.49~$\pm$~0.04 \\ 
G333.093$+$01.966 & 1.6~$\pm$~0.6 & 7.67~$\pm$~0.30 & 7.0~$\pm$~0.3 & 11.1~$\pm$~1.3 & 21.7~$\pm$~1.4 & 18.2$_{-2.6}^{+2.5}$ & 3.5$_{-2.5}^{+2.2}$ & 6.4$_{-3.1}^{+3.0}$ & 4.2$_{-2.4}^{+2.2}$ & 29.0 & 0.7$_{-0.1}^{+0.1}$ & 2.2$_{-1.8}^{+1.6}$ & 0.48~$\pm$~0.06 \\ 
G333.129$-$00.439 & 3.5~$\pm$~0.4 & 7.09~$\pm$~0.30 & 6.7~$\pm$~0.6 & 12.1~$\pm$~0.7 & 39.0~$\pm$~3.7 & 28.2$_{-9.8}^{+5.6}$ & 10.8$_{-5.7}^{+9.9}$ & 15.7$_{-10.1}^{+4.8}$ & 11.2$_{-4.7}^{+10.0}$ & $>$99 & 1.2$_{-0.4}^{+0.2}$ & 3.1$_{-1.7}^{+2.8}$ & 0.55~$\pm$~0.05 \\ 
G333.215$-$00.298 & 3.4~$\pm$~0.4 & 7.11~$\pm$~0.30 & 8.4~$\pm$~0.5 & 12.1~$\pm$~0.7 & 39.0~$\pm$~3.7 & 23.7$_{-9.1}^{+7.7}$ & 15.3$_{-8.4}^{+8.2}$ & 12.9$_{-9.3}^{+7.8}$ & 14.0$_{-8.0}^{+8.2}$ & 18.7 & 1.0$_{-0.4}^{+0.3}$ & 4.5$_{-2.5}^{+2.5}$ & 0.47~$\pm$~0.04 \\ 
G333.245$-$00.487 & 3.4~$\pm$~0.4 & 7.11~$\pm$~0.30 & 7.6~$\pm$~1.0 & 12.0~$\pm$~0.8 & 40.0~$\pm$~3.5 & 30.2$_{-13.8}^{+4.9}$ & 9.8$_{-4.6}^{+12.1}$ & 18.7$_{-13.7}^{+5.7}$ & 9.3$_{-5.0}^{+12.5}$ & 16.4 & 1.3$_{-0.6}^{+0.2}$ & 2.9$_{-1.4}^{+3.6}$ & 0.50~$\pm$~0.06 \\ 
G333.593$-$00.090 & 3.2~$\pm$~0.4 & 7.17~$\pm$~0.30 & 7.1~$\pm$~0.4 & 15.3~$\pm$~2.1 & 39.7~$\pm$~1.2 & 29.1$_{-5.4}^{+5.7}$ & 10.6$_{-5.2}^{+5.7}$ & 13.2$_{-5.8}^{+5.5}$ & 11.2$_{-5.8}^{+5.2}$ & $>$99 & 1.2$_{-0.2}^{+0.2}$ & 3.3$_{-1.7}^{+1.8}$ & 0.68~$\pm$~0.10 \\ 
G333.594$-$00.291 & 3.2~$\pm$~0.4 & 7.17~$\pm$~0.30 & 6.2~$\pm$~0.7 & 12.7~$\pm$~0.5 & 41.2~$\pm$~1.2 & 26.8$_{-6.7}^{+7.6}$ & 14.5$_{-8.4}^{+6.3}$ & 12.5$_{-6.4}^{+8.9}$ & 15.9$_{-8.4}^{+6.2}$ & 14.8 & 1.1$_{-0.3}^{+0.3}$ & 4.5$_{-2.7}^{+2.1}$ & 0.60~$\pm$~0.04 \\ 
G333.627$-$00.199 & 3.2~$\pm$~0.4 & 7.16~$\pm$~0.30 & 6.2~$\pm$~0.8 & 14.7~$\pm$~1.8 & 40.0~$\pm$~1.1 & 25.8$_{-3.9}^{+7.1}$ & 14.1$_{-6.0}^{+4.1}$ & 9.8$_{-4.5}^{+6.9}$ & 15.3$_{-6.1}^{+3.9}$ & 20.3 & 1.1$_{-0.2}^{+0.3}$ & 4.4$_{-1.9}^{+1.4}$ & 0.69~$\pm$~0.09 \\ 
G333.683$-$00.512 & 11.8~$\pm$~0.4 & 7.10~$\pm$~0.30 & 8.5~$\pm$~0.2 & 12.1~$\pm$~0.7 & 39.0~$\pm$~3.7 & 24.0$_{-8.2}^{+8.5}$ & 15.0$_{-9.2}^{+8.3}$ & 13.3$_{-8.7}^{+8.6}$ & 13.6$_{-8.7}^{+8.6}$ & 32.5 & 1.6$_{-0.5}^{+0.5}$ & 1.3$_{-0.8}^{+0.7}$ & 0.72~$\pm$~0.06 \\ 
G336.446$-$00.198 & 10.3~$\pm$~0.4 & 6.55~$\pm$~0.30 & 12.5~$\pm$~1.1 & 14.3~$\pm$~1.0 & 41.6~$\pm$~1.4 & 22.9$_{-9.7}^{+6.8}$ & 18.7$_{-6.3}^{+9.4}$ & 14.6$_{-9.3}^{+6.1}$ & 12.8$_{-6.3}^{+9.6}$ & 55.7 & 1.3$_{-0.6}^{+0.4}$ & 1.8$_{-0.6}^{+0.9}$ & 0.50~$\pm$~0.10 \\ 
G336.455$+$00.042 & 11.4~$\pm$~0.4 & 6.84~$\pm$~0.30 & 13.0~$\pm$~0.5 & 14.3~$\pm$~1.0 & 41.6~$\pm$~1.4 & 19.4$_{-6.3}^{+10.1}$ & 22.2$_{-9.7}^{+6.6}$ & 11.0$_{-6.4}^{+10.9}$ & 16.2$_{-9.7}^{+6.6}$ & $>$99 & 1.2$_{-0.4}^{+0.6}$ & 1.9$_{-0.9}^{+0.6}$ & 0.51~$\pm$~0.08 \\ 
G337.957$-$00.474 & 3.1~$\pm$~0.4 & 7.14~$\pm$~0.30 & 11.7~$\pm$~1.0 & 17.1~$\pm$~1.3 & 48.7~$\pm$~2.1 & 26.6$_{-7.6}^{+9.5}$ & 21.5$_{-9.7}^{+7.9}$ & 14.3$_{-7.9}^{+9.0}$ & 17.3$_{-10.9}^{+7.7}$ & 30.4 & 1.1$_{-0.3}^{+0.4}$ & 6.9$_{-3.2}^{+2.7}$ & 0.54~$\pm$~0.07 \\ 
G338.706$+$00.645 & 4.3~$\pm$~0.4 & 6.76~$\pm$~0.30 & 9.5~$\pm$~0.3 & 14.0~$\pm$~0.7 & 32.6~$\pm$~1.4 & 26.3$_{-9.8}^{+7.9}$ & 6.3$_{-1.9}^{+9.6}$ & 15.0$_{-10.3}^{+4.5}$ & 9.6$_{-5.6}^{+4.0}$ & 13.2 & 1.1$_{-0.4}^{+0.3}$ & 1.5$_{-0.5}^{+2.2}$ & 0.51~$\pm$~0.04 \\ 
G338.911$+$00.615 & 4.4~$\pm$~0.4 & 6.73~$\pm$~0.30 & 8.5~$\pm$~0.7 & 14.1~$\pm$~0.7 & 33.2~$\pm$~1.5 & 22.4$_{-6.0}^{+5.9}$ & 10.8$_{-5.9}^{+5.6}$ & 9.9$_{-5.8}^{+5.9}$ & 9.2$_{-5.9}^{+5.6}$ & $>$99 & 1.0$_{-0.3}^{+0.3}$ & 2.4$_{-1.4}^{+1.3}$ & 0.56~$\pm$~0.05 \\ 
G338.934$-$00.067 & 3.2~$\pm$~0.4 & 7.10~$\pm$~0.30 & 11.3~$\pm$~0.5 & 17.8~$\pm$~1.5 & 45.2~$\pm$~1.7 & 26.8$_{-7.4}^{+7.8}$ & 18.3$_{-7.8}^{+7.0}$ & 13.3$_{-7.7}^{+7.6}$ & 14.1$_{-7.4}^{+6.8}$ & 25.4 & 1.1$_{-0.3}^{+0.3}$ & 5.7$_{-2.5}^{+2.3}$ & 0.59~$\pm$~0.07 \\ 
G339.233$+$00.243 & 11.2~$\pm$~0.4 & 6.64~$\pm$~0.30 & 12.9~$\pm$~0.3 & 17.0~$\pm$~0.8 & 43.3~$\pm$~1.3 & 22.9$_{-6.7}^{+8.7}$ & 20.1$_{-7.9}^{+7.4}$ & 12.1$_{-6.8}^{+7.9}$ & 13.7$_{-7.2}^{+7.2}$ & 76.8 & 1.4$_{-0.4}^{+0.5}$ & 1.8$_{-0.7}^{+0.7}$ & 0.68~$\pm$~0.06 \\ 
G340.051$-$00.231 & 3.9~$\pm$~0.4 & 6.86~$\pm$~0.30 & 8.9~$\pm$~0.3 & 14.5~$\pm$~0.7 & 36.0~$\pm$~0.6 & 18.5$_{-2.2}^{+10.0}$ & 17.5$_{-9.9}^{+1.9}$ & 6.0$_{-2.6}^{+8.8}$ & 15.5$_{-8.6}^{+2.3}$ & 5.2 & 0.8$_{-0.1}^{+0.4}$ & 4.5$_{-2.6}^{+0.7}$ & 0.55~$\pm$~0.04 \\ 
G340.294$-$00.193 & 3.5~$\pm$~0.4 & 6.98~$\pm$~0.30 & 10.5~$\pm$~1.0 & 14.6~$\pm$~0.7 & 36.1~$\pm$~0.6 & 21.9$_{-4.4}^{+6.9}$ & 14.1$_{-6.8}^{+4.6}$ & 6.2$_{-1.9}^{+11.3}$ & 15.2$_{-11.5}^{+4.6}$ & $>$99 & 0.9$_{-0.2}^{+0.3}$ & 4.0$_{-2.0}^{+1.4}$ & 0.48~$\pm$~0.06 \\ 
G341.090$-$00.017 & 3.4~$\pm$~0.7 & 4.30~$\pm$~0.40 & 8.5~$\pm$~0.9 & 17.0~$\pm$~1.5 & 36.4~$\pm$~0.3 & 20.9$_{-4.8}^{+6.6}$ & 15.5$_{-6.9}^{+4.6}$ & 7.6$_{-4.7}^{+6.7}$ & 11.6$_{-6.5}^{+4.7}$ & 29.8 & 0.9$_{-0.2}^{+0.3}$ & 4.6$_{-2.2}^{+1.6}$ & 0.56~$\pm$~0.08 \\ 
G341.283$-$00.353 & 3.4~$\pm$~0.7 & 6.00~$\pm$~0.60 & 9.3~$\pm$~0.4 & 17.0~$\pm$~1.5 & 36.4~$\pm$~0.3 & 23.8$_{-4.9}^{+4.3}$ & 12.4$_{-4.1}^{+4.9}$ & 5.1$_{-1.5}^{+9.4}$ & 9.4$_{-4.2}^{+4.6}$ & 16.0 & 1.0$_{-0.2}^{+0.2}$ & 3.7$_{-1.4}^{+1.6}$ & 0.60~$\pm$~0.08 \\ 
G345.235$+$01.408 & 2.4~$\pm$~0.5 & 6.00~$\pm$~0.60 & 4.2~$\pm$~0.9 & 9.9~$\pm$~2.2 & 25.4~$\pm$~2.7 & 22.1$_{-6.5}^{+6.6}$ & 3.3$_{-1.0}^{+6.2}$ & 8.9$_{-5.1}^{+3.6}$ & 6.7$_{-2.1}^{+4.7}$ & 57.4 & 0.9$_{-0.3}^{+0.3}$ & 1.4$_{-0.5}^{+2.6}$ & 0.48~$\pm$~0.10 \\ 
G345.410$-$00.953 & 2.6~$\pm$~0.6 & 6.96~$\pm$~0.00 & 6.4~$\pm$~0.5 & 12.4~$\pm$~0.6 & 31.0~$\pm$~1.0 & 24.3$_{-5.3}^{+3.0}$ & 6.8$_{-2.6}^{+5.4}$ & 9.6$_{-3.6}^{+4.5}$ & 9.1$_{-4.2}^{+3.9}$ & $>$99 & 1.0$_{-0.2}^{+0.1}$ & 2.6$_{-1.2}^{+2.1}$ & 0.54~$\pm$~0.04 \\ 
G345.491$+$00.354 & 2.0~$\pm$~0.4 & 7.20~$\pm$~0.70 & 10.0~$\pm$~0.3 & 14.8~$\pm$~0.7 & 34.2~$\pm$~0.5 & 17.4$_{-5.2}^{+9.6}$ & 16.8$_{-9.6}^{+5.0}$ & 5.3$_{-0.5}^{+9.2}$ & 14.0$_{-9.2}^{+4.2}$ & 4.2 & 0.7$_{-0.2}^{+0.4}$ & 8.4$_{-5.1}^{+3.0}$ & 0.48~$\pm$~0.04 \\ 
G348.249$-$00.971 & 2.5~$\pm$~0.8 & 7.27~$\pm$~0.30 & 7.6~$\pm$~0.2 & 13.4~$\pm$~1.1 & 35.7~$\pm$~3.2 & 23.6$_{-5.4}^{+8.1}$ & 12.0$_{-7.4}^{+5.0}$ & 10.7$_{-5.3}^{+7.5}$ & 11.6$_{-7.3}^{+5.0}$ & $>$99 & 0.9$_{-0.2}^{+0.3}$ & 4.8$_{-3.3}^{+2.5}$ & 0.53~$\pm$~0.05 \\ 
G348.261$+$00.485 & 1.3~$\pm$~0.3 & 7.30~$\pm$~0.70 & 8.1~$\pm$~1.8 & 13.6~$\pm$~1.1 & 36.5~$\pm$~3.0 & 28.9$_{-8.6}^{+3.9}$ & 7.6$_{-2.6}^{+9.3}$ & 16.2$_{-9.3}^{+3.8}$ & 6.8$_{-3.1}^{+9.0}$ & 28.8 & 1.1$_{-0.3}^{+0.1}$ & 5.9$_{-2.4}^{+7.3}$ & 0.50~$\pm$~0.09 \\ 
G348.691$-$00.826 & 3.4~$\pm$~0.3 & 6.93~$\pm$~0.30 & 3.4~$\pm$~0.9 & 8.0~$\pm$~0.8 & 27.4~$\pm$~2.0 & 20.1$_{-4.0}^{+3.4}$ & 7.6$_{-3.9}^{+2.9}$ & 8.4$_{-4.3}^{+3.7}$ & 11.5$_{-4.0}^{+2.7}$ & 84.3 & 0.8$_{-0.2}^{+0.1}$ & 2.2$_{-1.2}^{+0.9}$ & 0.48~$\pm$~0.06 \\ 
G348.710$-$01.044 & 3.4~$\pm$~0.3 & 7.15~$\pm$~0.10 & 3.3~$\pm$~0.8 & 8.0~$\pm$~0.8 & 27.4~$\pm$~2.0 & 23.5$_{-6.4}^{+7.0}$ & 3.9$_{-0.6}^{+6.3}$ & 11.7$_{-6.5}^{+3.5}$ & 7.7$_{-2.3}^{+6.5}$ & 20.3 & 0.9$_{-0.3}^{+0.3}$ & 1.1$_{-0.2}^{+1.9}$ & 0.50~$\pm$~0.05 \\ 
G349.814$-$00.625 & 2.2~$\pm$~0.4 & 6.30~$\pm$~0.60 & 3.6~$\pm$~0.9 & 9.9~$\pm$~0.5 & 29.8~$\pm$~1.0 & 20.2$_{-3.6}^{+5.2}$ & 9.3$_{-4.6}^{+4.0}$ & 7.1$_{-3.5}^{+5.8}$ & 12.7$_{-5.6}^{+3.5}$ & 58.9 & 0.8$_{-0.1}^{+0.2}$ & 4.2$_{-2.2}^{+2.0}$ & 0.51~$\pm$~0.05 \\ 
G350.617$+$00.984 & 1.4~$\pm$~0.3 & 9.00~$\pm$~0.90 & 5.9~$\pm$~1.1 & 10.5~$\pm$~1.3 & 27.0~$\pm$~2.5 & 21.8$_{-3.8}^{+4.0}$ & 5.3$_{-2.4}^{+2.8}$ & 7.9$_{-4.0}^{+4.7}$ & 8.9$_{-3.2}^{+2.6}$ & 5.9 & 0.8$_{-0.1}^{+0.1}$ & 3.8$_{-1.9}^{+2.2}$ & 0.52~$\pm$~0.08 \\ 
G350.995$+$00.654 & 1.4~$\pm$~0.3 & 7.67~$\pm$~0.30 & 4.2~$\pm$~2.1 & 12.1~$\pm$~1.0 & 31.8~$\pm$~2.1 & 24.3$_{-3.4}^{+4.5}$ & 6.9$_{-3.5}^{+3.5}$ & 8.1$_{-3.6}^{+5.4}$ & 11.1$_{-4.3}^{+3.5}$ & 81.4 & 0.9$_{-0.1}^{+0.2}$ & 4.9$_{-2.7}^{+2.7}$ & 0.61~$\pm$~0.10 \\ 
G351.130$+$00.449 & 1.6~$\pm$~1.3 & 6.65~$\pm$~0.10 & 5.4~$\pm$~2.5 & 12.1~$\pm$~1.0 & 31.9~$\pm$~2.1 & 26.8$_{-8.6}^{+8.0}$ & 8.8$_{-3.4}^{+4.8}$ & 12.9$_{-9.8}^{+1.3}$ & 6.9$_{-1.7}^{+9.3}$ & 29.0 & 1.0$_{-0.3}^{+0.3}$ & 5.5$_{-5.0}^{+5.4}$ & 0.53~$\pm$~0.12 \\ 
G351.170$+$00.704 & 17.1~$\pm$~2.1 & 5.61~$\pm$~0.00 & 5.9~$\pm$~1.6 & 2.9~$\pm$~1.9 & 46.5~$\pm$~1.7 & 23.9$_{-10.9}^{+13.0}$ & 23.3$_{-13.6}^{+9.8}$ & 20.8$_{-10.6}^{+11.9}$ & 22.6$_{-13.5}^{+10.4}$ & $>$99 & 2.1$_{-1.0}^{+1.2}$ & 1.4$_{-0.8}^{+0.6}$ & 0.20~$\pm$~0.25 \\ 
G351.246$+$00.673 & 1.3~$\pm$~0.1 & 8.56~$\pm$~0.10 & 4.1~$\pm$~1.5 & 11.9~$\pm$~1.2 & 31.3~$\pm$~2.7 & 24.8$_{-7.4}^{+4.6}$ & 6.5$_{-4.2}^{+2.1}$ & 8.5$_{-2.7}^{+5.1}$ & 11.0$_{-4.1}^{+3.2}$ & 17.1 & 0.9$_{-0.3}^{+0.2}$ & 5.0$_{-3.3}^{+1.7}$ & 0.64~$\pm$~0.08 \\ 
G351.311$+$00.663 & 1.3~$\pm$~0.1 & 7.71~$\pm$~0.30 & 2.6~$\pm$~0.3 & 11.0~$\pm$~0.9 & 34.2~$\pm$~2.5 & 23.4$_{-7.0}^{+8.6}$ & 10.7$_{-7.7}^{+3.2}$ & 7.3$_{-1.2}^{+8.3}$ & 15.9$_{-7.6}^{+4.8}$ & 28.7 & 0.9$_{-0.3}^{+0.3}$ & 8.2$_{-5.9}^{+2.6}$ & 0.63~$\pm$~0.04 \\ 
G351.348$+$00.593 & 1.4~$\pm$~0.3 & 7.67~$\pm$~0.30 & 4.4~$\pm$~1.9 & 11.0~$\pm$~0.9 & 34.2~$\pm$~2.5 & 24.8$_{-3.6}^{+5.7}$ & 9.4$_{-5.4}^{+4.1}$ & 10.5$_{-4.9}^{+6.1}$ & 12.6$_{-6.1}^{+4.5}$ & 54.2 & 0.9$_{-0.1}^{+0.2}$ & 6.7$_{-4.1}^{+3.2}$ & 0.56~$\pm$~0.09 \\ 
G351.367$+$00.640 & 1.3~$\pm$~0.1 & 9.70~$\pm$~0.10 & 4.5~$\pm$~1.4 & 11.9~$\pm$~1.2 & 31.3~$\pm$~2.7 & 25.9$_{-3.4}^{+4.4}$ & 5.4$_{-3.6}^{+2.8}$ & 8.8$_{-4.2}^{+3.7}$ & 10.6$_{-3.3}^{+3.2}$ & 22.7 & 1.0$_{-0.1}^{+0.2}$ & 4.2$_{-2.8}^{+2.2}$ & 0.66~$\pm$~0.08 \\ 
G351.383$+$00.737 & 1.3~$\pm$~0.1 & 7.71~$\pm$~0.30 & 2.1~$\pm$~1.6 & 11.1~$\pm$~0.7 & 31.9~$\pm$~1.7 & 25.5$_{-3.6}^{+3.0}$ & 3.2$_{-1.0}^{+6.4}$ & 12.1$_{-7.3}^{+3.6}$ & 12.3$_{-3.2}^{+3.5}$ & 27.2 & 0.9$_{-0.1}^{+0.1}$ & 2.5$_{-0.8}^{+4.9}$ & 0.65~$\pm$~0.07 \\ 
G351.472$-$00.458 & 3.5~$\pm$~0.7 & 7.46~$\pm$~0.10 & 6.2~$\pm$~0.7 & 10.9~$\pm$~0.7 & 28.0~$\pm$~1.0 & 21.2$_{-5.8}^{+3.5}$ & 6.8$_{-3.6}^{+5.6}$ & 9.1$_{-5.8}^{+3.7}$ & 8.0$_{-3.4}^{+5.8}$ & 11.1 & 0.9$_{-0.2}^{+0.1}$ & 1.9$_{-1.1}^{+1.6}$ & 0.51~$\pm$~0.05 \\ 
G351.651$+$00.510 & 1.4~$\pm$~0.3 & 5.60~$\pm$~0.60 & 9.2~$\pm$~1.1 & 14.2~$\pm$~0.5 & 44.0~$\pm$~0.9 & 32.7$_{-15.6}^{+9.8}$ & 11.3$_{-3.4}^{+15.5}$ & 22.1$_{-15.7}^{+6.6}$ & 7.7$_{-1.1}^{+15.4}$ & 50.8 & 1.2$_{-0.6}^{+0.4}$ & 8.1$_{-3.0}^{+11.2}$ & 0.41~$\pm$~0.05 \\ 
G351.688$-$01.169 & 14.5~$\pm$~1.1 & 7.56~$\pm$~0.10 & 2.6~$\pm$~0.7 & 8.0~$\pm$~1.0 & 25.1~$\pm$~1.5 & 22.2$_{-4.9}^{+6.7}$ & 2.9$_{-1.2}^{+4.2}$ & 9.2$_{-5.2}^{+2.8}$ & 7.9$_{-2.4}^{+4.8}$ & 34.4 & 1.6$_{-0.4}^{+0.5}$ & 0.2$_{-0.1}^{+0.3}$ & 0.98~$\pm$~0.13 \\ 
G352.597$-$00.188 & 6.6~$\pm$~0.3 & 7.56~$\pm$~0.20 & 10.2~$\pm$~0.6 & 14.0~$\pm$~0.6 & 36.5~$\pm$~0.6 & 26.2$_{-9.6}^{+7.9}$ & 10.3$_{-1.7}^{+9.5}$ & 14.8$_{-9.8}^{+4.4}$ & 7.7$_{-0.8}^{+9.8}$ & 10.5 & 1.2$_{-0.4}^{+0.4}$ & 1.6$_{-0.3}^{+1.4}$ & 0.54~$\pm$~0.05 \\ 
G353.038$+$00.581 & 1.8~$\pm$~0.4 & 6.40~$\pm$~0.60 & 4.2~$\pm$~2.3 & 15.9~$\pm$~1.2 & 47.3~$\pm$~4.2 & 33.4$_{-9.5}^{+6.7}$ & 13.9$_{-6.6}^{+9.0}$ & 15.3$_{-10.1}^{+6.1}$ & 16.1$_{-6.6}^{+8.9}$ & 56.6 & 1.3$_{-0.4}^{+0.3}$ & 7.7$_{-4.1}^{+5.3}$ & 0.73~$\pm$~0.11 \\ 
G353.076$+$00.287 & 1.8~$\pm$~0.4 & 5.00~$\pm$~0.50 & 3.6~$\pm$~1.4 & 16.1~$\pm$~1.2 & 46.7~$\pm$~3.5 & 32.3$_{-8.5}^{+6.9}$ & 14.5$_{-7.2}^{+7.6}$ & 13.5$_{-7.3}^{+8.9}$ & 17.0$_{-9.2}^{+6.5}$ & $>$99 & 1.2$_{-0.3}^{+0.3}$ & 8.0$_{-4.4}^{+4.6}$ & 0.72~$\pm$~0.08 \\ 
G353.092$+$00.857 & 1.8~$\pm$~0.4 & 5.50~$\pm$~0.60 & 2.3~$\pm$~1.0 & 14.6~$\pm$~1.7 & 35.9~$\pm$~4.1 & 32.0$_{-10.1}^{+9.6}$ & 3.9$_{-0.4}^{+9.5}$ & 14.2$_{-10.2}^{+4.3}$ & 7.2$_{-2.2}^{+9.6}$ & 14.1 & 1.2$_{-0.4}^{+0.4}$ & 2.2$_{-0.5}^{+5.3}$ & 0.72~$\pm$~0.08 \\ 
G353.408$-$00.381 & 3.3~$\pm$~0.9 & 8.48~$\pm$~0.10 & 12.0~$\pm$~0.5 & 13.4~$\pm$~0.5 & 43.7~$\pm$~1.4 & 27.2$_{-9.7}^{+6.4}$ & 16.5$_{-5.2}^{+9.1}$ & 17.3$_{-9.9}^{+5.8}$ & 12.9$_{-5.2}^{+9.4}$ & 93.2 & 1.1$_{-0.4}^{+0.3}$ & 5.0$_{-2.1}^{+3.1}$ & 0.40~$\pm$~0.03 \\ 
G359.740$-$00.412 & 1.5~$\pm$~0.3 & 8.60~$\pm$~0.90 & 13.4~$\pm$~2.8 & 33.9~$\pm$~3.4 & 80.6~$\pm$~18.0 & 70.1$_{-35.6}^{+21.0}$ & 26.1$_{-10.3}^{+17.6}$ & 41.0$_{-35.0}^{+12.3}$ & 22.2$_{-9.7}^{+15.4}$ & 21.4 & 2.6$_{-1.3}^{+0.8}$ & 17.4$_{-7.7}^{+12.2}$ & 1.17~$\pm$~0.19 \\ 

\hline
\end{longtable}}
\end{landscape}
\end{center}

% Don't change these lines
\bsp	% typesetting comment
\label{lastpage}
\end{document}